\documentclass[aps,pre,onecolumn,superscriptaddress,amsmath,amssymb]{revtex4}

\usepackage[utf8]{inputenc}
\usepackage{amsmath,amsthm}
\usepackage{graphicx}
\usepackage[colorlinks = true,
            linkcolor = blue,
            urlcolor  = blue,
            citecolor = blue,
            anchorcolor = blue]{hyperref}
\usepackage{braket}
\usepackage{xcolor}
\newcommand{\be}{\begin{equation}}
\newcommand{\ee}{\end{equation}}
\newcommand{\bea}{\begin{eqnarray}}
\newcommand{\eea}{\end{eqnarray}}

\newcommand{\one}{\mathbb{I}}

\global\long\def\l{\la}

\global\long\def\ell#1{\theta_{#1}}
\global\long\def\bell#1{\tilde\theta_{#1}}

\global\long\def\la{\lambda} \global\long\def\ka{\kappa}
\global\long\def\si{\sigma}

\global\long\def\s{\sigma}

\global\long\def\eps{\epsilon}
\global\long\def\al{\alpha}

\def\XYZ{XY\! Z}
\def\XXZ{XXZ}

\global\long\def\no{\nonumber}

\theoremstyle{thm@}

\theoremstyle{remark}

\def\multiplet{single particle subspace}

\def\ir{{\mathrm i}}

\begin{document}

\title{Invariant subspaces and explicit Bethe vectors in the integrable open
  spin $1/2$ $\XYZ$ chain }

\author{Xin  Zhang}
\affiliation{Beijing National Laboratory for Condensed Matter Physics, Institute of Physics, Chinese Academy of Sciences, Beijing 100190, China}
\author{  Andreas Kl\"umper} 
 \affiliation{Department of Physics,
  University of Wuppertal, Gaussstra\ss e 20, 42119 Wuppertal,
  Germany}
\author{Vladislav Popkov}
\affiliation{Faculty of Mathematics and Physics, University of Ljubljana, Jadranska 19, SI-1000 Ljubljana, Slovenia}
 \affiliation{Department of Physics,
  University of Wuppertal, Gaussstra\ss e 20, 42119 Wuppertal,
  Germany}

\begin{abstract}
We derive a criterion under which splitting of all eigenstates of an open
$\XYZ$ Hamiltonian with boundary fields into two invariant subspaces,
spanned by chiral shock states, occurs. The splitting is governed by an
integer number, which has the geometrical meaning of the maximal number of
kinks in the basis states. We describe the generic structure of the respective
Bethe vectors. We obtain explicit expressions for Bethe vectors,  in the
  absence of Bethe roots, and those generated by one Bethe root, and
investigate the \multiplet.  We also describe in detail an elliptic analogue
of the spin-helix state, appearing in both the periodic and the open $\XYZ$ model, and
derive the eigenstate condition.  The elliptic analogue of the spin-helix
state is characterized by a quasi-periodic modulation of the magnetization
profile, governed by Jacobi elliptic functions.
\end{abstract}
\maketitle

\section{Introduction}

Exact solutions are indispensable for our understanding of statistical
mechanics of interacting systems \cite{BaxterBook,Korepin}.  The paradigmatic
spin-$\frac12$ $\XYZ$ chain is one of the most fascinating models in quantum
statistical mechanics \cite{BaxterBook}.  Its 2D classical statistical
counterpart, the 8-vertex model, has appeared as the first example
possessing continuously varying critical exponents.  The periodic $\XYZ$ spin
chain with an even number of sites was solved by Baxter \cite{BaxterBook}.
Takhtajan and Faddeev recovered Baxter’s solution via the generalized
algebraic Bethe ansatz method \cite{Takhtajan}. For open systems, several
approaches have been proposed to construct integrable structures
\cite{Hou1993,DeVega1994} and exact solutions
\cite{Fan1996,Yang2006,Faldella2014}.  Bethe Ansatz equations for the spectrum
of the $\XYZ$ model with generic integrable boundary conditions (including
periodic, anti-periodic and open boundary conditions) were first derived by
the off-diagonal Bethe ansatz method \cite{Cao2013,Cao2014,OffDiagonal}.

Despite many years of studies, little is known about the structure of the $\XYZ$
eigenstates (Bethe vectors), especially for an open system.

It is our purpose to show that on special manifolds of parameters, the $\XYZ$
eigenstate problem can be significantly advanced, and we are able to
explicitly link the solutions of the Bethe ansatz equations (BAE) to the
coefficients of the Bethe vectors in a special chiral basis. To this end, we
prove a splitting of the whole Hilbert space into two subspaces invariant with
respect to the action of the open $\XYZ$ Hamiltonian with tuned boundary
fields. Along the way, we derive a criterion of such a splitting to occur.
For the simplest cases, we calculate explicitly the Bethe vectors for
arbitrary system size and unveil their geometrical intepretation.

The analogous Hilbert space splitting in the $\XXZ$ open spin chain has been
proven with phantom Bethe roots \cite{PhantomLong,PhantomBetheAnsatz}. Thus,
our results can be viewed as a generalization of the phantom $\XXZ$ Bethe states
concept onto the fully anisotropic $\XYZ$ spin chain case.

As for the $\XXZ$ model, we refer to our finding as the ``splitting theorem''
which gives us a tool to study the structure of Bethe states in the $\XYZ$
model, belonging to each invariant subspace. We name the set of Bethe states
belonging to an invariant subspace a multiplet. The number of independent
states in the multiplet is equal to the dimension of the respective invariant
subspace and is given by $dim(M)=\binom{N}{0}+ \binom{N}{1}+ \ldots +
\binom{N}{M}$, where $M$ is a nonnegative integer, ranging from $0$ to $N-1$.
Choosing $M$ fixes the manifold in the parameter space. There are actually
  several disconnected submanifolds corresponding to the same $M$,
  parametrized by another integer $L_0$, see criterion
  (\ref{SplittingCriterion}).

Each individual state in the multiplet can be parametrized by exactly $M$
Bethe roots.  The simplest multiplet ($M=0$) has just one state in it and this
state is an elliptic analogue of the spin-helix state \cite{PhantomShort,
  2017SHS-Linbdlad-Gunter,2017SHS-Carlo}.  The next simplest case corresponds
to the cases $M=1, 2$ and so forth. Then the multiplet already contains a
large, polynomially growing with system size number of
states. Investigating this large set of states for large systems can be used
for statistical analysis. Here we treat in detail the case $M=1$, for
which the smallest multiplet containing $N+1$ states.  We find the explicit form
of Bethe vectors and use them to calculate various observables.

The plan of the manuspript is as follows: After introducing the model we
derive a local divergence condition which appears crucial for our study. 
On the base of it, we describe the simplest possible eigenstate, the elliptic analogue of the 
spin-helix state.  
Next, we formulate the criterion (\ref{SplittingCriterion}), under which a  splitting of
the Hilbert space into two invariant chiral subspaces occurs, and describe the basis
states spanning the invariant subspaces.  In the final part of the manuscript we use the gained
knowledge to investigate the elliptic analogue of phantom Bethe states belonging to the invariant
subspace with dimension $N+1$ where $N$ is the length of the $\XXZ$ spin
chain. Details of the proofs are given in the Appendix.

\section{Factorized elliptic spin helix  eigenstates in $\XYZ$ spin chain}

The Hamiltonian of the $\XYZ$ spin chain with generic open boundaries is
\begin{align}
&H=\sum_{n=1}^{N-1}{h}_{n,n+1}+\vec{h}_{1} \vec{\si}_1+\vec{h}_{N} \vec{\si}_N,\label{Hamiltonian}\\
&{h}_{n,n+1}=J_x{\sigma}_n^x\sigma_{n+1}^x+J_y\sigma_n^y\sigma_{n+1}^y+J_z\sigma_n^z\sigma_{n+1}^z,\label{bulk}\\
&\vec{h}_{1} \vec{\si}_1=h_x^-\sigma_1^x+h_y^-\sigma_1^y+h_z^-\sigma_1^z, \label{BC;1}\\
&\vec{h}_{N} \vec{\si}_N=h_x^+\s_N^x+h_y^+\s_N^y+h_z^+\s_N^z.\label{BC;2}
\end{align}
where the anisotropy parameter and the boundary magnetic fields are parameterized
as in \cite{Yang2006,OffDiagonal} \footnote{Here we rewrite the expressions
  after some simple transformations.}
\begin{align}
&J_x=\frac{\ell{4}(\eta)}{\ell{4}(0)},\quad J_y=\frac{\ell{3}(\eta)}{\ell{3}(0)},\quad J_z=\frac{\ell{2}(\eta)}{\ell{2}(0)},\label{Jxyzparameterization}\\
&h_z^{\mp}=\mp\frac{\ell{1}(\eta)}{\ell{2}(0)}\prod_{l=1}^3\frac{\ell{2}(\alpha_l^\pm)}{\ell{1}(\alpha_l^\mp)}\\
&h_x^{\mp}=\pm\frac{\ell{1}(\eta)}{\ell{4}(0)}\prod_{l=1}^3\frac{\ell{4}(\alpha_l^\mp)}{\ell{1}(\alpha_l^\mp)},\\
&h_y^{\mp}=\mp i\frac{\ell{1}(\eta)}{\ell{3}(0)}\prod_{l=1}^3\frac{\ell{3}(\alpha_l^\mp)}{\ell{1}(\alpha_l^\mp)}.
\end{align}
where we use the following shorthand notations $ \ell{\al},\,\bell{\al}$ 
\begin{align}
	\ell{\al}(u) \equiv  \vartheta_{\al} (\pi u,e^{i\pi\tau}),\quad \bell{\al}(u) \equiv   \vartheta_{\al} (\pi u ,e^{2i\pi\tau}),\quad {\rm Im}[\tau]>0,\quad \alpha=1,2,3,4,
\end{align}
for the elliptic Jacobi theta functions $\vartheta_{\al} (\pi u,e^{i\pi\tau})$
given for instance in Ref.~\cite{WatsonBook}.\\

\textbf{Hermiticity condition:} Hermiticity of the Hamiltonian in the bulk is
guaranteed by choosing $\eta, {\rm i}\tau\in \mathbb R$.  With this choice,
and with an appropriate rotation of the axes, any set of spin anisotropies
($J_x,  J_y,  J_z$) can be realized.  For hermiticity of the boundary fields we
have to demand the following restrictions for the boundary parameters $\al_k^\pm$:
\begin{align}
	&\mathrm{Im}[\alpha_1^+]=\mathrm{Im}[\alpha_1^-]=0,\quad \mathrm{Im}[\alpha_2^+]=\tfrac{(2k_1+1)\tau}{2i},\quad
	\mathrm{Im}[\alpha_2^-]=\tfrac{(2k_2+1)\tau}{2i},\no\\
	&\mathrm{Re}[\alpha_3^+]=\tfrac{2k_3+1}{2},\quad\mathrm{Re}[\alpha_3^-]=\tfrac{2k_4+1}{2},\qquad k_1,k_2,k_3,k_4\in\mathbb{Z}.
\end{align}

Our analysis starts by the following remarkable observation. Define a local
state \cite{Fan1996}
\begin{align}
\psi(u)&=\binom{\bell{1}(u) }{-\bell{4}(u)},\label{def;psi}
\end{align}
where $u\in\mathbb C$ is a free parameter. The vector $\psi(u)$ satisfies the
following divergence condition \cite{MPA2021}
\bea
&&h\, \left[ \psi(u)\otimes\psi(u+\eta) \right]=\left[a(u)\sigma^z\otimes \one_2- a(u+\eta) \one_2 \otimes \sigma^z + d(u) \one_4\right]\psi(u)\otimes\psi(u+\eta),\label{eq:divcond1}\eea
where $h=\sum_\alpha J_\alpha \sigma^\alpha\otimes \sigma^\alpha$ is a
$4\times 4$ Hamiltonian density operator, and the functions $a(u),\,d(u)$ are
given by (see Appendix \ref{Proof;Div}):
\bea
&&a(u) =  \frac{\ell{1}(\eta) \ell{2}(u)}{  \ell{2}(0) \ell{1}(u) },\quad d(u)= g(\eta)+g(u)-g(u+\eta),\quad g(u) = \frac{\ell{1}(\eta)\ell{1}'(u)}{\ell{1}'(0) \ell{1}(u)}. \label{a(u,s)}
\eea

Using Eq.~(\ref{eq:divcond1}), we find a family of spatially inhomogeneous
factorizable eigenstates of the $XY\! Z$ model with boundary fields:
\bea
&&\left[\sum_{n=1}^{N-1}h_{n,n+1}-a(u_1) \sigma^z_1+a(u_N)\sigma^z_N\right]\ket{\Psi_+} = E\ket{\Psi_+}, \label{eq:XYZa} \\ 
&&\ket{\Psi_+} = \bigotimes_{n=1}^N \psi(u_n),\qquad E=\sum_{n=1}^{N-1} d(u_n),\quad u_n=u_1+(n-1)\eta. \label{eq:sep}
\eea

Noticing that the Hamiltonian is invariant under $\eta \to-\eta $, we can
write another divergence condition, complementary to (\ref{eq:divcond1}) 
\bea
&&h\,\psi(u)\otimes\psi(u-\eta)=\left[-a(u)\sigma^z\otimes \one_2 + a(u-\eta) \one_2 \otimes \sigma^z + d(-u)\one_4\right]\psi(u)\otimes\psi(u-\eta),
\label{eq:divcond2}
\eea
rendering the state $\ket{\Psi_{-}} = \bigotimes_{n=1}^{N} \psi(u_{2-n})$
 an eigenvector for the Hamiltonian $\sum_{n=1}^{N-1}h_{n,n+1}+a(u_{1})
 \sigma^z_1-a(u_{2-N})\sigma^z_N$.
 
\begin{figure}[htbp]
\centerline{\includegraphics[width=0.42\textwidth]{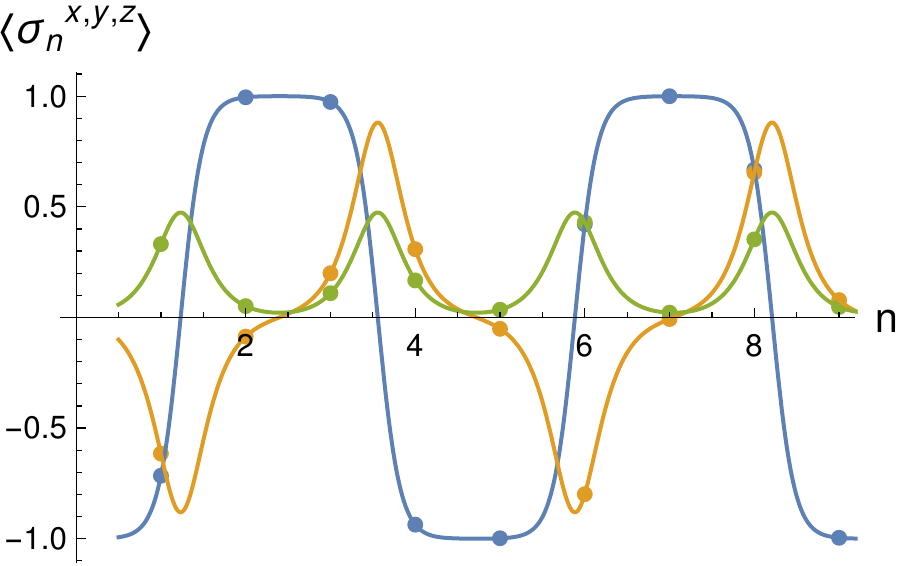}}\vspace{5mm}
\caption{Local magnetization profiles for the factorized eigenstate
  $\ket{\Psi_{+}}$ from (\ref{eq:sep}) discussed in detail in section
  \ref{sec:M=0}. The $x, y, z$-components of the local magnetization are given
  by blue, yellow and green points, respectively. The used parameters are:
  $\eta=0.43, \tau=0.35 i$, $u_1=0.9 + 0.23 i$, $N=9$. The curves are exact
  interpolations by elliptic functions (\ref{sn})-(\ref{dn}) and are guides
  for the eye.}
\label{Fig-SHSell}
\end{figure}

Remarkably, both states $\ket{\Psi_+}$ from (\ref{eq:sep}) and the state
$\ket{\Psi_{-}}$ correspond to periodic modulations of the local magnetization
components as exemplified in Fig.~\ref{Fig-SHSell}.  Note that the individual
  qubit states are all pure, and correspondingly, all spins are fully
  polarized, $\langle\si_n^x \rangle^2 + \langle\si_n^y \rangle^2
  +\langle\si_n^z \rangle^2=1$ for all $n$.  

Since the magnetization profile of the state (\ref{eq:sep}) is given by
elliptic functions, we will call this state an elliptic spin-helix state
  (elliptic SHS), an elliptic generalization of the spin-helix states of the
$\XXZ$ model which is governed by trigonometric functions \cite{PhantomShort,
  2017SHS-Linbdlad-Gunter,2017SHS-Carlo}. 
  Further physical properties of the elliptic SHS, e.g. the projections of the
  magnetization vector lying on conic sections are given in section \ref{sec:M=0}.

The elliptic SHS (\ref{eq:sep}) visualized in Fig.~\ref{Fig-SHSell} is the
simplest nontrivial state of an open $\XYZ$ Hamiltonian and it is remarkable
in many respects. First, due to the local divergence property
(\ref{eq:divcond2}) it is also an eigenstate of the periodic system
$H=\sum_{n=1}^N h_{n,n+1}$, with $\vec{\sigma}_{N+1} \equiv \vec{\sigma}_1$, if periodicity conditions $a(u_1)
= a(u_{N+1})$ are fulfilled, which is equivalent to
 \begin{align}
N \eta = 2 L_0 \tau + 2 K_0, \quad
L_0,K_0\in\mathbb{Z} \label{Condition:ellSHS-periodic}
\end{align}
 Note that for the periodic quantum chain, $u_1$ in (\ref{eq:sep}) can be
 chosen arbitrarily leading to a multiplet of factorized states with the same
 energy 
 $$E=\sum_{n=1}^N d(u_n)\equiv Ng(\eta).$$

For the Hermitian case, we let $L_0=0$ and $\eta=2K_0/N$.
The existence of a remarkably simple elliptic SHS Fig.~\ref{Fig-SHSell}
  in a periodic Hermitian $\XYZ$ spin chain is related in the case of even length
  to a special arrangement of Bethe roots $\la_j$ in the form of a perfect
  equidistant string, parallel to the real axis,
    \begin{align}
\la_{j+1}  = \la_j +\frac{1}{M}, \quad j=1,2, \ldots ,M-1,  
  \label{eq:ellSHS-BetheString}
\end{align}
where $M=N/2$ is an integer.
The $\XYZ$ elliptic SHS string (\ref{eq:ellSHS-BetheString}) is analogous to the
string of phantom Bethe roots appearing in the $\XXZ$ model
\cite{PhantomShort}. Other linearly independent states can be obtained by
shifting the phases $u_n \rightarrow u_n + \beta $ in (\ref{eq:XYZa}). We can show
that the degeneracy of elliptic SHS
in the periodic chain satisfying (\ref{Condition:ellSHS-periodic}) is $deg=
2N$. More details will be given elsewhere.

Notably, the elliptic SHS (\ref{eq:sep}) is a quasi-stationary state in
\textit{any} open systems with $\XYZ$ bulk dynamics.  Indeed, the action
of the open Hamiltonian on it
\begin{align}
&\left(\sum_{n=1}^{N-1}h_{n,n+1}- E \right)\ket{\Psi_+}  = \left[ a(u_1) \sigma^z_1-a(u_N)\sigma^z_N\right]\ket{\Psi_+} 
\end{align}
only affects the state at the boundaries, while the bulk stays intact
before the information from the boundaries will spread over all the
system, requiring a time of the order $N/v_c$ where $v_c$ is the sound
velocity. This quasi-stationarity makes the elliptic SHS long-lived states,
thus making them attractive for e.g.~cold atom experiments where the
anisotropies can be tuned and individual spins can be manipulated,
\cite{JepsenPRX,Jepsen2021}.

Third, the elliptic SHS can be generated also dissipatively, in open $\XYZ$
spin chains with boundary dissipation projecting the boundary spins onto
predefined qubit states. Namely, if the first spin is projected onto the pure
qubit state $\psi(u_1)$ and the last spin $N$ onto the state $\psi(u_1+(N-1)
\eta)$, then the interior spins will relax towards the elliptic SHS
(\ref{eq:sep}) with time, provided that the dissipation is sufficiently
strong, see \cite{MPA2021}.

Analogously to (\ref{def;psi}), we introduce the following bra vector
\begin{align}
\phi(u)=\left(\bell{1}(u),\,-\bell{4}(u)\right)\equiv[\psi(u)]^t.\label{phi}
\end{align}
As $h^{t}=h$, two divergence conditions for $\phi(u)\equiv[\psi(u)]^t$ can be
obtained from (\ref{eq:divcond1}) and (\ref{eq:divcond2}) by transposition.
More properties about $\psi(u)$ and $\phi(u)$ can be found in Appendix
\ref{Proof;ab}.

\section{Invariant subspaces for $\XYZ$ Hamiltonian}\label{sec:Theorem}
 
The properties (\ref{eq:divcond1}), (\ref{eq:divcond2}) and (\ref{h;phi_1})-
(\ref{h;phi_2}) are fundamental. They entail the splitting of all eigenstates
of the open anisotropic $\XYZ$ Heisenberg Hamiltonian on special manifolds
into two complementary chiral invariant subspaces. These manifolds are
characterized by model parameters satisfying the following criterion, see
Appendix:
\begin{align}
(N-1-2M)\eta=\sum_{\sigma=\pm}\sum_{k=1}^3\epsilon_k^\sigma\al_k^\sigma-2L_0\tau+2K_0,\quad \prod_{\sigma=\pm}\prod_{k=1}^3\epsilon_k^\sigma=-1,\quad \epsilon_k^\sigma=\pm 1, \quad  L_0,K_0\in\mathbb{Z},\label{SplittingCriterion}
\end{align}
where $M$ is an integer which takes values $0\leq M\leq N-1$. For the generic
case ($N-1-2M\neq 0$), once certain boundary parameters $\{\al_l^\pm\}$ are
selected, $\eta$ should take certain discrete values in the complex
plane. Noticing that
\begin{align}
H|_{\eta\to\eta+2} =H,\quad H|_{\eta\to\eta+2\tau} =e^{-4i\pi(u+\tau)}H,
\end{align}
the integers $L_0$ and $K_0$ in (\ref{SplittingCriterion}) can be restricted
to $0\leq L_0,K_0<N-1-2M$.

Eq.~(\ref{SplittingCriterion}) has an important symmetry, namely the
substitution
\begin{align}
&M \rightarrow N-1-M, \quad \epsilon_k^\sigma \rightarrow-\epsilon_k^\sigma, \quad L_0,K_0 \rightarrow -L_0,-K_0
\end{align}
leaves Eq.~(\ref{SplittingCriterion}) invariant.

\begin{figure}[htbp]
\centerline{
\includegraphics[width=0.5\textwidth]{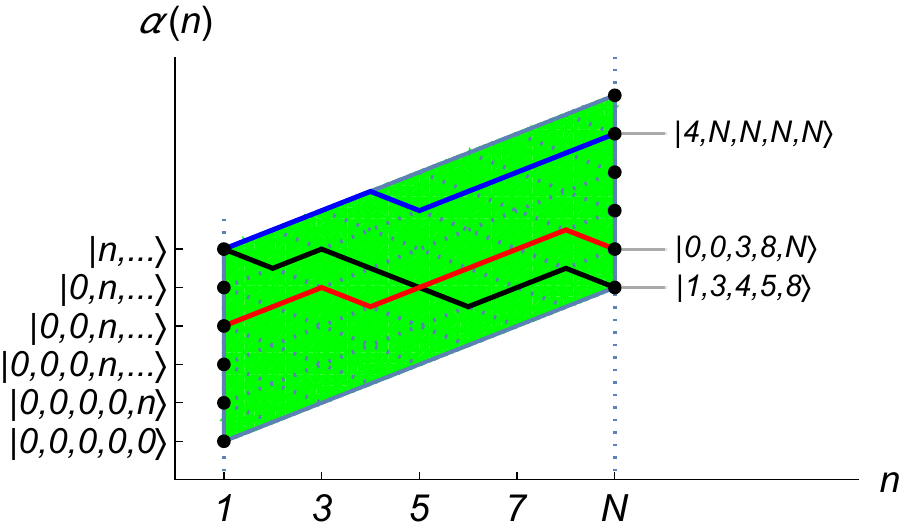}}
\centerline{
\includegraphics[width=0.5\textwidth]{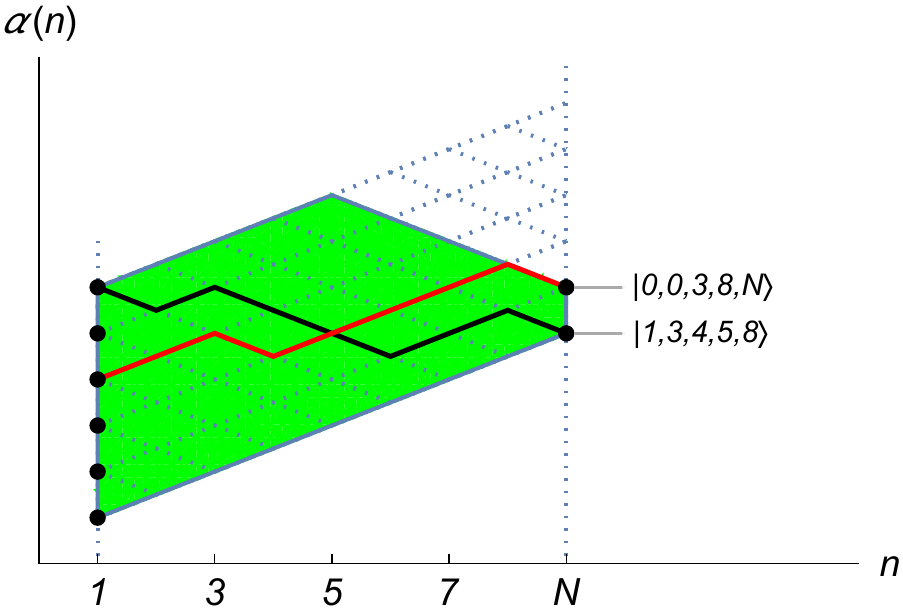}}
\caption{ \textbf{Upper Panel}: Symmetric set (not all states are linearly
  independent!) of all states in (\ref{KetVecs}) is represented by all
  possible trajectories (directed paths), lying within the filled green
  region, including the boundaries.  Each individual path starts in one of
  $M_{+}+1$ points (filled black circles) on site $n=1$ and ends at one of
  $M_{+}+1$ points at $n=N$.  \textbf{Lower Panel} shows the set of linearly
  independent states (\ref{Basis;Kets}), which is a subset of (\ref{KetVecs})
  in the upper panel. The allowed trajectories end in one of two points at the
  right boundary site $n=N$, their total number being $ \binom{N}{0}+
  \binom{N}{1}+ \ldots \binom{N}{M_{+}}$.  Black and red trajectories indicate
  two basis states from (\ref{Basis;Kets}): $\ket{1,3,4,5,8}$ and $\ket{0,0,3,8,N}$,
  respectively.  The blue trajectory in the upper panel represents a state
  $\ket{4,N,N,N,N}$ which can be expressed via basis states
  (\ref{Basis;Kets}). Note that $M_{+}$ gives the maximal number of kinks
    a trajectory can have.}
\label{Fig-Mgrid}
\end{figure}

Basis states of these subspaces are given by the factorized products of states
of type $\otimes_{n=1}^N\psi\!\left(u_{\al(n)}\right)$ with neighboring sites
parameters $\al(n),\al(n+1)$ satisfying the $\al(n+1)=\al(n)\pm 1 $
restriction.  This fact allows to represent the basis vectors geometrically,
plotting the ``phase" $\al(n)$ versus $n$, and joining the points $\al(n)$ by
a line. Then, each basis vector corresponds to a trajectory consisting of
segments with constant positive or negative slopes as shown in
Fig.~\ref{Fig-Mgrid}. Each such segment represents a piece of some factorized
state of type (\ref{eq:sep}).

Let the boundary parameters in $H$ satisfy the $\XYZ$ splitting criterion
(\ref{SplittingCriterion}) with $M_{+}=M$. Denote
\begin{align}
M_{-}=N-1-M_{+}.
\end{align}
Define two parameters $u_1$, $v_1$ by
\begin{align}
	&u_1=\frac{1}{2}-\sum_{k=1}^3\epsilon_k^-\al_k^-,\label{u0}\\
	&v_1=-u_1-\tau.\label{v0}
\end{align}
One invariant subspace  is spanned by the following ket-vectors
\begin{align}
&|\underbrace{0,\dots,0}_{m_0},\underbrace{n_1,\dots,n_{k}}_{k} ,\underbrace{N,\dots,N}_{m_N}\,\rangle\no\\
&=\bigotimes_{r_1=1}^{n_1}\psi(u_{r_1}-2m_0\eta)\bigotimes_{r_2=n_1+1}^{n_2}\psi(u_{r_2}-2(m_0+1)\eta)\dots\bigotimes_{r_{k+1}=n_k+1}^{N}\psi(u_{r_{k+1}}-2(m_0+k)\eta),\label{KetVecs}\\
&0<n_1<n_2<\cdots<n_k<N, \quad m_0,\,k,\,m_N\geq 0,\quad m_0+k+m_N=M_+,\quad \no
\end{align}
and another (complementary) invariant subspace is spanned by bra vectors 
\begin{align}
&\langle\!\langle\,\underbrace{0,\dots,0}_{m'_0},\underbrace{n'_1,\dots,n'_{k'}}_{k'} ,\underbrace{N,\dots,N}_{m'_N}|,\label{BraVecs}
\end{align}
obtained by replacing $m_0,\,k,\,m_N,\,\psi,\,u_1,\,M_+,\,\{n_1,\ldots,n_k\}$
in (\ref{KetVecs}) by
$m'_0,\,k',\,m'_N,\,\phi,\,v_1,\,M_-,\,\{n'_1,\ldots,n'_{k'}\}$ respectively,
with $M_-=N-M_{+}-1$, and a subsequent transposition.

For convenience, we rewrite the vectors in Eqs.~(\ref{KetVecs})-(\ref{BraVecs}) as
\begin{align}
|n_1,\ldots,n_{M_+}\rangle,\quad \hbox{and}\quad \langle\!\langle n'_1,\ldots,n'_{M_-}|.
\end{align}

Now we formulate our main statement regarding the splitting of the Hilbert space:

\bigskip

\textbf{Theorem} \textit{
The set of linearly independent ket states 
\begin{align}
|0,\ldots,0\rangle, \,\,|0,\ldots,0,n_1\rangle,\,\,|0,\ldots,0,n_1,n_2\rangle,\ldots,|n_1,\ldots,n_{M_+}\rangle,\quad 1\leq n_l\leq N,\quad n_l<n_{l+1}.\label{Basis;Kets}
\end{align} 
and the set of linearly independent bra states 
\begin{align}
\langle\!\langle 0,\ldots,0|,\, \langle\!\langle0,\ldots,0,n'_1|,\, \langle\!\langle 0,\ldots,0,n'_1,n'_2|,\ldots,  \langle\!\langle n'_1,\ldots,n'_{M_-}|,\quad 1\leq n'_l\leq N,\quad n'_l<n'_{l+1}.\label{Basis;Bras}
\end{align} 
form two bi-orthogonal complementary subspaces $G_M^{+}$ and $G_M^{-}$
invariant under the action of $H$ satisfying (\ref{SplittingCriterion}) with
$M_{+}:=M$ The dimensions are $dim\ G_M^{+}= \binom{N}{0}+ \binom{N}{1}+
\ldots \binom{N}{M_{+}}$ and $dim\ G_M^{-}=2^N- dim\ G_M^{+}$. The parameters
$u_1,v_1$ for the sets (\ref{Basis;Kets}), (\ref{Basis;Bras}) are given in
(\ref{u0}),(\ref{v0}). }

\bigskip

Thus, all the eigenvalues of $H$ satisfying (\ref{SplittingCriterion}) split
into two families: the right eigenvectors for the first family are given by
linear combinations of (\ref{Basis;Kets}) while the left eigenvectors for
the (complementary) second family are given by linear combinations of
(\ref{Basis;Bras}).

\textit{Remark}. One notices that the set of independent vectors in the
theorem is not symmetric: both coordinates $0$ and $N$ appear several
  times in (\ref{KetVecs}), but not in (\ref{Basis;Kets}). So not all vectors
in (\ref{KetVecs}) are included in (\ref{Basis;Kets}).  Indeed, the vectors
(\ref{KetVecs}) are not all linearly independent; the number of linearly
independent vectors is given by dimension $dim(G_M^+)$, the number of
vectors in (\ref{Basis;Kets}).  We visualize both the symmetric and the
minimal (linearly independent) sets in Fig.~\ref{Fig-Mgrid}.

The proof of the theorem follows that for the partially anisotropic $\XXZ$
model \cite{PhantomLong} and is given in Appendix \ref{Proof;Theorem}.

Next we will give the explicit form of the Bethe vectors satisfying the
splitting criterion (\ref{SplittingCriterion}), linking the coefficients in
the expansion to the solution of the Bethe Ansatz equations.

\section{Elliptic analogues of spin-helix eigenstates in the $\XYZ$ model }
\label{sec:M=0}

If we choose the boundary fields to satisfy our splitting criterion
(\ref{SplittingCriterion}) with $M_{+}=0$, then according to our theorem $H$
has a 1-dimensional invariant subspace $G_M^{+}$ consisting of just one state,
a fully factorized state $\ket{\Psi_+} = \bigotimes_{n=1}^N
\psi(u_n)$. Indeed, the boundary fields satisfy
\begin{align}
&\vec{h}_1\vec{\si}_1\psi(u_1)=-a(u_1) \si_1^z \psi(u_1)+b_L(u_1)\psi(u_1),\label{hL} \\
&\vec{h}_N\vec{\si}_N \psi(u_N)=a(u_N) \si_N^z \psi(u_N)+b_R(u_N)\,\psi(u_N), \label{hR} 
\end{align}
(see Eq.~(\ref{u0;Ket})) making the factorized state $\ket{\Psi_+}$ from
(\ref{eq:sep}) an eigenstate of the Hamiltonian with eigenvalue
$\sum_{n=1}^{N-1}b(u_n)+b_L(u_1)+b_R(u_N)$.  The explicit expressions of
$b_L(u), \ b_R(u)$ are given in the Appendix.

Since the eigenstate $\ket{\Psi_+}$ is factorized, it is fully described by
its one-point observables, i.e.~the components of the magnetization profile,
$\langle \si_n^\al \rangle$ with $\al=x,y,z$. We find
\begin{align}
	& \langle \si_n^+ \rangle
	=-\frac{\bell{1}^*(u_n)\bell{4}(u_n)}{|\bell{1}(u_n)|^2+|\bell{4}(u_n)|^2},\\
	& \langle \si_n^z \rangle=\frac{|\bell{1}(u_n)|^2-|\bell{4}(u_n)|^2 }{|\bell{1}(u_n)|^2+|\bell{4}(u_n)|^2},
\end{align}
where $\si_n^{+}= (\si_n^{x}+ i \si_n^{y})/2$, and $^*$ denotes complex
conjugation. Using $\bell{1}^* (u)= \bell{1} (u^*)$, $\bell{4}^*(u)=\bell{4}
(u^*)$, the identities (\ref{theta;11})-(\ref{theta;14}), and the relation
between the functions $\bell{\al} $ and the Jacobi elliptic functions ${\rm sn,
  cn, dn}$, and assuming $ \eta$ being real, we readily obtain
\begin{align}
	&\langle \si_n^x \rangle = 2 {\rm Re}[\langle \si_n^+ \rangle]=- \frac {\ell{1} (\beta_n) \ell{2} ( i\gamma) }   {\ell{4} (\beta_n) \ell{3} (i\gamma) } =A_x \,{\rm sn}(2 K_k\,\beta_n,k ) \label{sn} \\
	&\langle \si_n^y \rangle  = 2 {\rm Im}[\langle \si_n^+ \rangle]=- i \frac {\ell{2} (\beta_n) \ell{1} (i\gamma) }   {\ell{4} (\beta_n) \ell{3} (i\gamma) }  = A_y \,{\rm cn}(2 K_k\,\beta_n,k )\,, \label{cn}  \\
	&\langle \si_n^z \rangle =-  \frac {\ell{3} (\beta_n) \ell{4} (i\gamma) }   {\ell{4} (\beta_n) \ell{3} (i\gamma) } = A_z\, {\rm dn}(2 K_k\,\beta_n,k ), 
	\label{dn} 
\end{align}
where $\beta_n={\rm Re}[u_n]={\rm Re}[u_1]+(n-1)\eta$, $\gamma={\rm
  Im}[u_n]={\rm Im}[u_1]$, while the elliptic function modulus $k$, quarter
period $K_k$ and the coefficients $A_\al$ are given by
\begin{align}
	&k=\frac{\ell{2}^2 (0) }{\ell{3}^2 (0) }\,, \quad k'=\sqrt{1-k^2}= \frac{\ell{4}^2 (0) }{\ell{3}^2 (0) }, \quad K_k=\frac{\pi\ell{3}^2(0)}{2},\\
	&A_x=-\sqrt{k} \ \frac {\ell{2}(i\gamma)}   {\ell{3} (i\gamma)}, \quad A_y =-i \sqrt{\frac{k}{k'}} \  \frac { \ell{1} ( i\gamma)}{\ell{3} (i\gamma) }\,,\quad
	A_z =-\frac{1}{\sqrt{k'}}\  \frac { \ell{4} ( i\gamma)} {\ell{3} (i\gamma)}. 
\end{align}
Note that in our physical case (for real $\eta$) $0\leq k \leq 1$, and $A_\al$
are also all real.  The periods of the sn, cn and dn functions in
Eqs.~(\ref{sn})-(\ref{dn}) are given by $2/\eta$, and $1/\eta$, in lattice
units.

For the parameterization (\ref{Jxyzparameterization}) with generic anisotropy,
i.e. generic value of $\tau$ corresponding to $J_x\ge J_y\ge J_z$, we find the
following.  The projection of the magnetization vector onto the $xy$ plane
orbits an ellipse, the projection onto the $yz$ plane lies on a finite
sector of an ellipse and the projection onto the $zx$ plane lies on a
finite sector of a hyperbola. For the special case of the SHS parameter
$\gamma$ taking the value $-\ir\tau/2$ we have a circular motion in the $xy$
plane with radius 1 and $z$-component 0. For $\gamma=0$ ($\gamma= -\ir\tau$)
the $y$-component of the magnetization vector is 0, the $x$-component takes
values from an interval symmetric around 0 and the $z$-component takes negative
(positive) values including $-1$ ($+1$).

In the critical $\XXZ$ limit, $\tau\to+\ir\infty$ with $k=0$ and
$J_x=J_y>J_z$, the projection of the magnetization vector onto the $xy$ plane
orbits a circle with radius between 0 and 1, the $z$-component being
constant.  For the non-critical $\XXZ$ limit, $\tau\to 0$ with $k=1$ and
$J_x>J_y=J_z$, the projection of the magnetization vector onto the $yz$ plane
lies on straight lines, the projections onto the other planes lie on (sectors
of) ellipses. Note however, that for $\eta\not=0$ most points cluster at
$x$-component $\pm1$ with the $y$- and $z$-components being 0.

We summarize these findings
\be
\frac{\langle \si_n^x \rangle^2}{A_x^2}+\frac{\langle \si_n^y
    \rangle^2}{A_y^2}=1,\quad
\frac{\langle \si_n^x \rangle^2}{A_x^2/k^2}+\frac{\langle \si_n^z
    \rangle^2}{A_z^2}=1,\quad
\frac{\langle \si_n^z \rangle^2}{A_z^2}-\frac{\langle \si_n^y
    \rangle^2}{A_y^2/k^2}=1  -k^2,
\ee
where only $A_z$ takes independent values
\be
A_x^2=1-(1-k^2)A_z^2,\quad
A_y^2=1-A_z^2,\quad
A_z\in [-1,+1].
\ee
These relations are consistent with $\langle\si_n^x \rangle^2 + \langle\si_n^y \rangle^2
  +\langle\si_n^z \rangle^2=1$.

\section{Single-kink Bethe eigenstates in the open $\XYZ$ model }
Let us specify our general results for the simplest yet nontrivial case
$M_{+}=1$ with invariant subspace $ G_M^{+}$ containing only $N+1$ basis
vectors which we call $\ket{0}, \ket{1}, \ldots \ket{N}$.  The action of the
$\XYZ$ Hamiltonian on these states straightforwardly gives
\begin{align}
&H \ket{n}=[E_0(1)+X(n)]\ket{n}+2A_{-}(n) \ket{n-1} + 2A_{+}(n)\ket{n+1},\quad n=1,2,\ldots N-1,\label{H;n}\\
&H \ket{0}=[E_0(1)+ X_L]\ket{0}+2A_{L}\ket{1},\label{H;1}\\
&H \ket{N}=[E_0(1)+X_R]\ket{N}+2A_R\ket{N-1},\label{H;N}
\end{align} 
where the expressions of some functions in the above formulas are shown in
Appendix \ref{Proof;ansatz}.

We search for the Bethe vectors belonging to the invariant subspace in the general form 
\begin{align}
	|\Psi(\lambda)\rangle=\sum_{n=0}^NF_{n}(\lambda)\ket{n},\quad 
	\mbox{with}\quad H|\Psi(\la)\rangle=E(\la)|\Psi(\la)\rangle, \label{PsiAnsatz}
\end{align}
where $F_n$ are coefficients which depend on the complex parameter $\la$
parametrizing the energy $E(\la)$ like
\begin{align}
	&E(\la)=E_0(1)+E_B(\la),\label{M1;Energy}\\
	&E_B(\la)=2[g(\la-\tfrac{\eta}{2})-g(\la+\tfrac{\eta}{2})].\label{Eb}
\end{align}
Substituting Eq.~(\ref{PsiAnsatz}) into (\ref{H;n})-(\ref{H;N}), we obtain a
system of linear equations for $F_n(\l)$, namely
\begin{align}
&F_{n+1}(\l)A_-(n+1)+F_{n-1}(\l)A_+(n-1)=Y(\l,n)F_n(\l), \label{recursive;1}\\	
&F_{0}(\l)A_L+F_{2}(\l)A_-(2)=Y(\l,1)F_1(\l),\label{recursive;2}\\
&F_{N}(\l)A_R+F_{N-2}(\l)A_+(N-2)=Y(\l,N-1)F_{N-1}(\l),\label{recursive;3}\\
&F_{1}(\l)A_-(1)=Y_L(\l)F_0(\l),\label{recursive;4}\\
&F_{N-1}(\l)A_+(N-1)=Y_R (\l)F_N(\l),\label{recursive;5}
\end{align}
where 
\begin{align}
&2Y(\l,n)=E_B(\l)-X(n),\label{Y;1} \\
&2Y_L(\l)=E_B(\l)-X_L,\label{Y;l}\\
&2Y_R(\l)=E_B(\l)-X_R\label{Y;r}.
\end{align}

We propose the following ansatz,
\begin{align}
	&F_{n}(\l)=\ka_n\left[B_{+}\,U_{n}(\l)+B_{-}\,U_{n}(-\l)\right],\quad n=0,\ldots,N,\label{Ansatz;1}\\
	&\ka_0=\frac{A_+(0)}{A_L},\quad \ka_N=\frac{A_-(N)}{A_R}, \quad 
	\ka_1=\ka_2=\ldots=\ka_{N-1}=1,\label{Ansatz;2}\\
	&U_{n}(\l)=\left[\frac{\ell{1}(\l+\frac{\eta}{2})}{\ell{1}(\l-\frac{\eta}{2})}\right]^{n}\frac{\ell{2}(\l-u_{n}+\frac{\eta}{2})}{\ell{2}(u_{n-1})\ell{2}(u_{n})},\label{def;U}
\end{align}
where $\{B_{\pm}\}$ are two $\l$-dependent constants. The functions
$\left\{U_{n}(\l)\right\}$ satisfies the identity
\begin{align}
	&U_{n+1}(\pm\l)A_-(n+1)+U_{n-1}(\pm\l)A_+(n-1)=Y(\l,n)U_{n}(\pm\l),\label{U;property}
\end{align}
for arbitrary $\l,\, n$. As a consequence,
Eqs.~(\ref{recursive;1})-(\ref{recursive;3}) are satisfied automatically.
From Eqs.~(\ref{recursive;4}),(\ref{recursive;5}) we get
\begin{align}
	\frac{B_-}{B_+}&=-\frac{\kappa_0\,Y_L(\l)U_0(\l)-A_-(1)U_1(\l)}{\ka_0\,Y_L(\l)U_0(-\l)-A_-(1)U_1(-\l)}\no\\
	\frac{B_-}{B_+}&=-\frac{\ka_NY_R(\l)U_N(\l)-A_+(N-1)U_{N-1}(\l)}{\ka_NY_R(\l)U_N(-\l)-A_+(N-1)U_{N-1}(-\l)}.\label{BAE0}
\end{align}

The consistency condition of the above gives the Bethe ansatz equation
  (BAE) which determines the Bethe root $\l$
\begin{align}
\frac{\kappa_0\,Y_L(\l)U_0(\l)-A_-(1)U_1(\l)}{\ka_0\,Y_L(\l)U_0(-\l)-A_-(1)U_1(-\l)}=\frac{\ka_NY_R(\l)U_N(\l)-A_+(N-1)U_{N-1}(\l)}{\ka_NY_R(\l)U_N(-\l)-A_+(N-1)U_{N-1}(-\l)}.\label{BAE1}
\end{align}

Recall that $$u_1=\frac12-\sum_{k=1}^3\epsilon_k^-\al_k^-,\quad
u_{N-2M_+}=\frac12+\sum_{k=1}^3\epsilon_k^+\al_k^+-2L_0\tau+2K_0,\quad
\prod_{k=1}^3\epsilon_k^-=-1,\quad \prod_{k=1}^3\epsilon_k^+=1.$$ We simplify
the expressions of $\kappa_0$ and $\kappa_N$ by substituting Eqs.~(\ref{u0}),
(\ref{Apm}), (\ref{AL}), (\ref{AR}) into (\ref{Ansatz;2})
\begin{align}
&\kappa_0
=\frac{\ell{1}(\sum_k\epsilon_k^-\al_k^-+2\eta)\prod_l\ell{1}(\al_l^-)}{\ell{1}(\eta)\prod_{j<k}\ell{1}(\eps_j^-\al_j^-+\eps_k^-\al_k^-+\eta)},\\
&\kappa_N=-\frac{\ell{1}(\sum_k\eps_k^+\al_k^++2\eta)\prod_{l}\ell{1}(\al_l^+)}{\ell{1}(\eta)\prod_{j<k}\ell{1}(\eps_j^+\al_j^++\eps_k^+\al_k^++\eta)}.
\end{align}
Especially, when $\eta=-\eps_k^-\al^-_k,\,k=1,2,3$, it is straightforward to
get $\ka_0=1$. Analogously $\ka_N=1$ when $\eta=-\eps_k^+\al^+_k,\,k=1,2,3$.

Suppose that our Hamiltonian is hermitian, so its spectrum $E(\l)$ is real.
Analyzing the expression for (\ref{Eb}) we find that $E_B(z)$ is an even
elliptic function in the complex plane of $z$ with periods $1,\tau$:
$E_B(z)=E_B(-z)$, $E_B(z+1)=E_B(z+\tau)=E_B(z)$. Consequently, we can restrict
the elementary domain of $z$ to the rectangle in the complex plane with $0\leq
{\rm Re}[z] \leq \frac12$, $0\leq {\rm Im}[z] \leq
\frac{\tau}{2i}$. Moreover, requiring the energies $E_B(\la)$ to be real
forbids the $\la$ to lie \textit{inside} the rectangle. Thus, all physically
valid values of $\la$ must lie on the edges of the rectangle ${\rm Re}[\la]=
0, \frac12$ and ${\rm Im}[\la]= 0, \frac{\tau}{2i}$.

To check our predictions, we select $\eps_3^-=-1,\eps_k^-=1,\,k=1,2$,
$\eps_l^+=1,\,l=1,2,3$ and diagonalize the Hamiltonian inside the invariant
subspace $G_1^{+}$ for sufficiently large systems, using the system of
equations (\ref{H;n})-(\ref{H;N}).

\begin{figure}[tbp]
\centerline{
\includegraphics[width=0.5\textwidth]{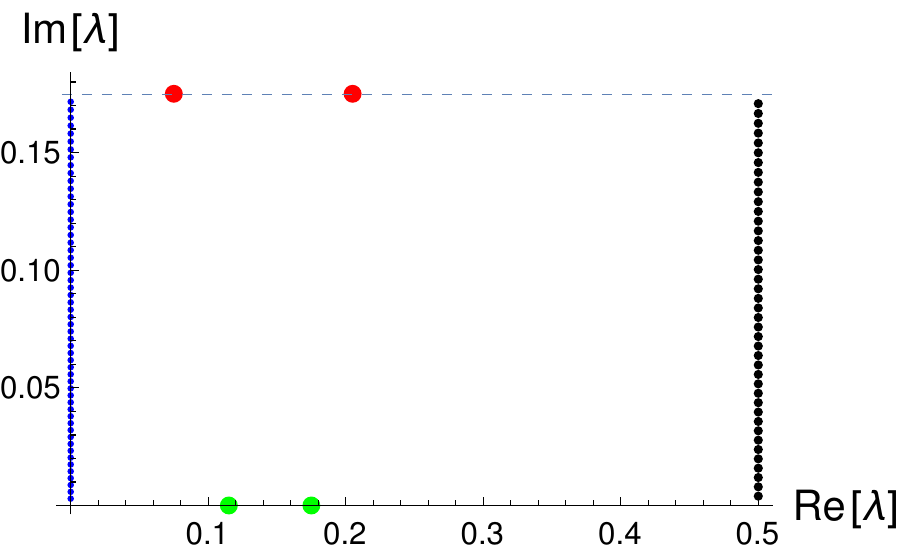}}
\centerline{
\includegraphics[width=0.5\textwidth]{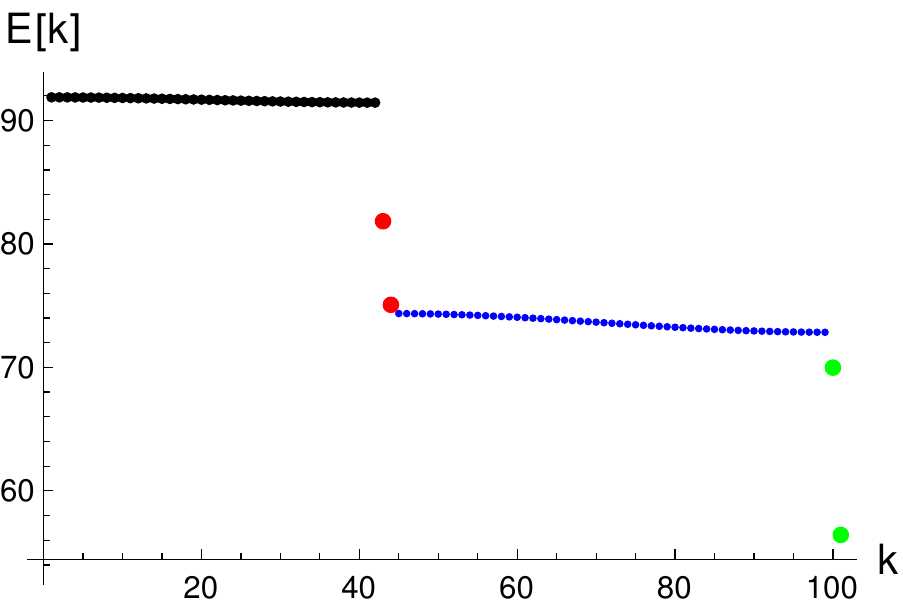}
}

\caption{\textbf{Upper panel}: Location of the Bethe roots in the multiplet
  $G_1^{+}$ with dimension $dim(G_1^{+})=N+1=101$. The respective energies
  decrease in the counter-clockwise direction, starting with the bottom of
    the ``continuum'' in the right.  The energies for the multiplet are shown
  in the \textbf{lower panel}, respecting the colour code.  Bulk parameters
  are: $N=100$, $\eta=0.43, \tau = 0.35 i$, and correspond to $J_x\approx
  4.51, J_y\approx 0.244, J_z\approx 0.136$.  Boundary parameters are
  $\{\al_{1}^-, \al_{2}^-,\al_{3}^-\}=\{0.9, 0.86 + \frac{\tau}{2}, \frac12 +
  0.53 i\}$, $\{\al_{1}^+, \al_{2}^+,\al_{3}^+\}= \{0.96, 38.99 + 1.225 i,
  \frac12 + 0.53 i\}$, and satisfy (\ref{SplittingCriterion}) with $L_0=2,
  \ K_0=0$ and $M_{+}=1$.  }
\label{Fig-BetheRoots}
\end{figure}

\begin{figure}[tbp]
\centerline{
\includegraphics[width=0.46\textwidth]{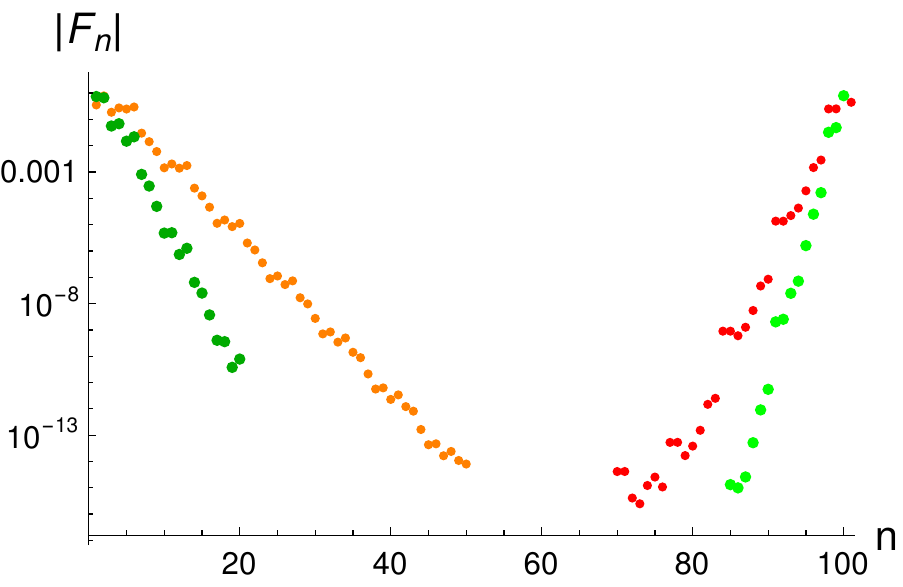}
}

\caption{Coefficients $|F_n|$ for the $H$ eigenfunctions corresponding to two
  red points and two green points in Fig.~\ref{Fig-BetheRoots} combined in one
  plot, on a logarithmic scale.  Red, orange, darkgreen, and green symbols
  corresponding to the energies $E=\ 81.844, \ 75.07, \ 70, \ 56.43$
  respectively.}
\label{Fig-LocalizedSol}
\end{figure}

The band structure of Eqs (\ref{H;n})-(\ref{H;N}) allows to easily solve the
problem, namely to obtain the coefficients of the Bethe vector $F_n$ and
the corresponding energies, as well as Bethe roots $\la_k$ for all Bethe
vectors belonging to the invariant subspace.  Some typical locations of the
Bethe roots inside $G_1^{+}$ is shown in Fig.~\ref{Fig-BetheRoots} for a
system of $N=100$ spins, satisfying our criterion (\ref{SplittingCriterion})
for $M_{+}=1$.  We see that most solutions correspond to either ${\rm
  Re}[\la_k]=0$ or ${\rm Re}[\la_k]=\frac12$, the roots being distributed
approximately homogeneously along the imaginary axis.  An inspection shows
that these roots correspond to the quasi-periodically changing coefficients
$F_n$, which are of order $O(1)$ for all $n$.  On the other hand, a few
separately located roots at the upper edge(${\rm
  Im}[\la_k]=\rm{Im}[\frac{\tau}{2}]$) and the lower edge (${\rm Im}[\la_k]=0$
(red and green in the Figure) correspond to the ``localized" cases, with
coefficients $F_n$ or $F_{N-n}$ decreasing exponentially, see
Fig.~\ref{Fig-LocalizedSol}. A similar type of solutions was also observed and
discussed in the context of $\XXZ$ model in \cite{PhantomLong}.

The origin of the ``localized" solutions in Fig.~\ref{Fig-BetheRoots} becomes
clear if we write down the BAE (\ref{BAE1}) in another, equivalent form (see
Appendix \ref{Proof;ansatz} for details),

\begin{align}
	&e^{-8i\pi L_0\l}\left[\frac{\ell{1}(\l+\frac{\eta}{2})}{\ell{1}(\l-\frac{\eta}{2})}\right]^{2N}\prod_{k=1}^3\frac{\ell{1}(\l-\al_k^+-\frac{\eta}{2})}{\ell{1}(\l+\al_k^++\frac{\eta}{2})}\frac{\ell{1}(\l+\al_3^--\frac{\eta}{2})}{\ell{1}(\l-\al_3^-+\frac{\eta}{2})}\prod_{l=1}^2\frac{\ell{1}(\l-\al_l^--\frac{\eta}{2})}{\ell{1}(\l+\al_l^-+\frac{\eta}{2})}=1.\label{BAE-M1equiv}
\end{align}
It is clear that solutions to (\ref{BAE-M1equiv}) with
$\left|\frac{\ell{1}(\l+\frac{\eta}{2})}{\ell{1}(\l-\frac{\eta}{2})}\right|=d\neq
1$ (meaning $2{\rm Re}[\la]\neq 0,1$) may lead to divergences for large $N$:
if $d>1$, the term $d^{2N}$ in (\ref{BAE-M1equiv}) diverges.  Therefore,
such a solution can only survive in the limit of $N \rightarrow \infty$,
if the divergence is compensated by one of factors in the numerator of
(\ref{BAE-M1equiv}) which becomes zero.  Due to $\ell{1}(0)=0$, the above
requirement amounts to one of arguments of $\ell{1}$ (\ref{BAE-M1equiv})
becoming zero, either $\la \rightarrow \al_{1,2,3}^{+} +\frac{\eta}{2} \ \ mod
\ 1$ or $\la\rightarrow \al_{1,2}^{-} + \frac{\eta}{2} \ \ mod \ 1$, or
$\la\rightarrow-\al_{3}^{-} + \frac{\eta}{2} \ \ mod \ 1$.  Indeed, in the
example shown in Fig.~\ref{Fig-BetheRoots}, we have four divergent solutions
featured in Fig.~\ref{Fig-LocalizedSol}; the corresponding Bethe roots, up to
exponentially small corrections are given by $\la
=\al_{1,2}^{\pm}+\frac{\eta}{2}$.

All the remaining $N-3$ solutions have the form ${\rm Re}(\la)=0$ or ${\rm
  Re}(\la)=\frac12$ corresponding to
$\left|\frac{\ell{1}(\l+\frac{\eta}{2})}{\ell{1}(\l-\frac{\eta}{2})}\right|=1$.
Their distribution on the segments $[0,\tau/2]$ of the imaginary axis becomes
approximately equidistant as $N$ grows. The upper and lower bounds for the
respective parts of the ``continuous spectrum" in the right panel of
Fig.~\ref{Fig-BetheRoots} are given by $E_0(1) + E_B(\frac12)$, $E_0(1) +
E_B(\frac{1+\tau}{2})$ for the black (upper) band and $E_0(1) +
E_B(\frac{\tau}{2})$, $E_0(1)+E_B(0)$ for the blue (lower) band.

\section*{Discussion}
Summarizing, we have proven a splitting of the Hilbert space into invariant
manifolds for the $\XYZ$ spin chain with boundary fields, and derived the spitting
condition (\ref{SplittingCriterion}). Our main result is intrinsically based
on a pair of local divergence conditions (\ref{eq:divcond1}), (\ref{eq:divcond2})
and is given as the Theorem in section \ref{sec:Theorem}.  Using our theorem, we
have demonstrated how one describes in detail the spectrum and
the structure of Bethe vectors within the invariant subspaces with $M_{+}=0$
and $M_{+}=1$ (one Bethe root). Generalizations of our results for arbitrary
$M_{+}$ is more technical and will be given elsewhere.

We find that our criterion (\ref{SplittingCriterion}) and our Bethe ansatz
equations are the same as the ones given in \cite{Yang2006}, which proves the
consistency of our results. In the limit $\tau\to +i\infty$, the $\XYZ$ chain
degenerates into the critical $\XXZ$ chain and we see the spin-helix structure
\cite{PhantomShort,MPA2021} in the Bethe vectors. The splitting criterion for
the $\XXZ$ chain has been derived in \cite{PhantomLong}.
	
It would be interesting to explore the consequences of the existence of the
chiral basis for the periodic $\XYZ$ chain.  We are convinced that our chiral
basis will be useful for the periodic system, since some results in
\cite{BaxterBook,Cao2013,OffDiagonal} show the existence of the homogeneous
BAE under certain conditions.  As a first step, we derived a condition for a
periodic system to have an elliptic SHS in (\ref{Condition:ellSHS-periodic}).

Another perspective seems to be opening in the cold atom experiments which
proved the possibility to create and sustain spin-helix states
\cite{JepsenPRX,Jepsen2021}, which are the simplest examples of so-called
phantom Bethe states \cite{PhantomShort} appearing in the $\XXZ$ Heisenberg
spin chain.  It would be very exciting if our novel elliptic SHS can be
prepared experimentally.

\begin{acknowledgments}
We acknowledge financial supports by the European Research Council through the
advanced grant No. 694544—OMNES (VP), and by the Deutsche
Forschungsgemeinschaft through DFG project KL 645/20-1 (VP). X.Z. thanks the
Institute of Physics, Chinese Academy of Sciences for financial support.
\end{acknowledgments}


\appendix
\section{Theta-functions and their properties}

In this paper, we adopt the notations of elliptic theta functions $\ell{\al}(u,q)$ following Ref. \cite{WatsonBook}
\begin{align}
\begin{aligned}
&\vartheta_{1}(u,q)=2\sum_{n=0}^\infty(-1)^n q^{(n+\frac12)^2}\sin[(2n+1)u],\\
&\vartheta_{2}(u,q)=2\sum_{n=0}^\infty q^{(n+\frac12)^2}\cos[(2n+1)u],\\
&\vartheta_{3}(u,q)=1+2\sum_{n=1}^\infty q^{n^2}\cos(2nu),\\
&\vartheta_{4}(u,q)=1+2\sum_{n=1}^\infty (-1)^mq^{n^2}\cos(2nu).
\end{aligned}
\end{align}
Recall that 
$\ell{\al}(u) \equiv  \vartheta_{\al} (\pi u,e^{i\pi\tau}),\,\, \bell{\al}(u) \equiv   \vartheta_{\al} (\pi u ,e^{2i\pi\tau})$.
There exist many identities for $\ell{\alpha}(u)$ and $\bell{\alpha}(u)$. Here
we list some useful identities
\cite{WatsonBook,OffDiagonal,2013Derkachev,MPA2021}
\begin{align}
	&\ell{2}(u)=\ell{1}(u+\tfrac12),\quad \ell{3}(u)=e^{i \pi(u+\frac{\tau }{4})}\ell{1}(u+\tfrac{1+\tau}{2}),\quad \ell{4}(u)=-e^{i\pi(u+\frac{\tau}{4}+\frac12)}\ell{1}(u+\tfrac{\tau}{2}),\label{four;theta}\\
	&\ell{1}(-u)=-\ell{1}(u),\quad \ell{1}(u+1)=-\ell{1}(u),\quad \ell{1}(u+\tau)=-e^{-2i\pi(u+\frac{\tau}{2})}\ell{1}(u),\label{periodicity}\\
	&\frac{\bell{1}(2u)}{\bell{4}(0)}=\frac{\ell{1}(u)\ell{2}(u)}{\ell{3}(0)\ell{4}(0)},\quad \frac{\bell{4}(2u)}{\bell{4}(0)}=\frac{\ell{3}(u)\ell{4}(u)}{\ell{3}(0)\ell{4}(0)},\label{Landen}\\
	&\ell{1}(2u)\ell{2}(0)\ell{3}(0)\ell{4}(0)=2\ell{1}(u)\ell{2}(u)\ell{3}(u)\ell{4}(u),\label{double;angle}\\
	&\bell{1}(u+v)\bell{1}(u-v)\bell{4}^2(0)=\bell{1}^2(u)\bell{4}^2(v)-\bell{1}^2(v)\bell{4}^2(u),\label{th_th;1}\\
	&\bell{4}(u+v)\bell{4}(u-v)\bell{4}^2(0)=\bell{4}^2(u)\bell{4}^2(v)-\bell{1}^2(v)\bell{1}^2(u),\label{th_th;2}\\
	&\ell{1}(u\pm v)\ell{2}(u\mp v)\ell{3}(0)\ell{4}(0)=\ell{1}(u)\ell{2}(u)\ell{3}(v)\ell{4}(v)\pm\ell{1}(v)\ell{2}(v)\ell{3}(u)\ell{4}(u).\label{addition;formula}\\
	&2\bell{1}(u+v)\bell{1}(u-v)=\ell{4}(u)\ell{3}(v)-\ell{4}(v)\ell{3}(u),\label{theta;11}\\
	&2\bell{4}(u+v)\bell{4}(u-v)=\ell{4}(u)\ell{3}(v)+\ell{4}(v)\ell{3}(u),\label{theta;44}\\
	&2\bell{4}(u+v)\bell{1}(u-v)=\ell{1}(u)\ell{2}(v)+\ell{1}(v)\ell{2}(u),\label{theta;14}\\
	&2\ell{1}(w)\ell{1}(x)\ell{1}(y)\ell{1}(z)\no\\
	&=\ell{1}(w')\ell{1}(x')\ell{1}(y')\ell{1}(z')+ \ell{2}(w')\ell{2}(x')\ell{2}(y')\ell{2}(z')\no\\
	&\quad-\ell{3}(w')\ell{3}(x')\ell{3}(y')\ell{3}(z')+\ell{4}(w')\ell{4}(x')\ell{4}(y')\ell{4}(z'),\label{additional_formula}\\
	&2\ell{4}(w)\ell{4}(x)\ell{4}(y)\ell{4}(z)\no\\
	&=\ell{1}(w')\ell{1}(x')\ell{1}(y')\ell{1}(z')- \ell{2}(w')\ell{2}(x')\ell{2}(y')\ell{2}(z')\no\\
	&\quad +\ell{3}(w')\ell{3}(x')\ell{3}(y')\ell{3}(z')+\ell{4}(w')\ell{4}(x')\ell{4}(y')\ell{4}(z'),\label{additional_formula;2}\\
	&o(u+x)o(u-x)o(v+y)o(v-y)-o(u+y)o(u-y)o(v+x)o(v-x)\no\\&=o(u+v)o(u-v)o(x+y)o(x-y),\quad o=\ell{1},\,\bell{1},\label{Riemann}
\end{align}
where 
\begin{align}
	&2w'=-w+x+y+z,\qquad 2x'=w-x+y+z,\no\\ &2y'=w+x-y+z,\qquad 2z'=w+x+y-z.\no
\end{align}
Define
\begin{align}
	\zeta(u)=\frac{\ell{1}'(u)}{\ell{1}(u)},\quad \tilde\zeta(u)=\frac{\bell{1}'(u)}{\bell{1}(u)}.
\end{align}
which possess the following properties 
\begin{align}
	&\zeta(u)=-\zeta(-u),\quad 
	\zeta(u+1)=\zeta(u),\quad \zeta(u+\tau)=\zeta(u)-2i\pi,\label{zeta;1}\\
	&\tilde\zeta(u)=-\tilde\zeta(-u),\quad 
	\tilde\zeta(u+1)=\tilde\zeta(u),\quad \tilde\zeta(u+2\tau)=\tilde\zeta(u)-2i\pi,\label{zeta;2}\\
	&2\tilde\zeta(2u)=\zeta(u)+\zeta(u+\tfrac12),\quad \zeta(u)=i\pi+\tilde\zeta(u)+\tilde\zeta(u+\tau),\label{zeta;3}\\
	&2\zeta(u)=2i\pi+\zeta(\tfrac{u}{2})+\zeta(\tfrac{u+1}{2})+\zeta(\tfrac{u+\tau}{2})+\zeta(\tfrac{u+\tau+1}{2}).\label{zeta;4}
\end{align}
The functions $\ell{\alpha}(u),\,\bell{\alpha}(u),\,\zeta(u),\,\bar\zeta(u)$
satisfy identities, e.g.
\begin{align}
	&\frac{\ell{2}(u)}{\ell{1}(u)}=\frac{\ell{2}(0)}{\ell{1}'(0)}\left[\zeta(\tfrac{u}{2})+\zeta(\tfrac{u+1}{2})-\zeta(u)\right],\label{zeta;sigma;1}\\
	&\frac{\ell{4}(u)}{\ell{1}(u)}=\frac{\ell{4}(0)}{\ell{1}'(0)}\left[\zeta(\tfrac{u}{2})+\zeta(\tfrac{u+\tau}{2})-\zeta(u)+i\pi\right],\label{zeta;sigma;2}\\
	&\frac{\ell{3}(u)}{\ell{1}(u)}=\frac{\ell{3}(0)}{\ell{1}'(0)}\left[\zeta(\tfrac{u}{2})+\zeta(\tfrac{u+1+\tau}{2})-\zeta(u)+i\pi\right],\label{zeta;sigma;3}\\
	&\frac{\ell{1}(x_1+x_2)\ell{1}(x_1+x_3)\ell{1}(x_2+x_3)}{\ell{1}(x_1)\ell{1}(x_2)\ell{1}(x_3)\ell{1}(x_1+x_2+x_3)}=\frac{1}{\ell{1}'(0)}\left[\zeta(x_1)+\zeta(x_2)+\zeta(x_3)-\zeta(x_1+x_2+x_3)\right],\label{zeta;sigma;4}\\
	&\frac{\bell{1}(x_1+x_2)\bell{1}(x_1+x_3)\bell{1}(x_2+x_3)}{\bell{1}(x_1)\bell{1}(x_2)\bell{1}(x_3)\bell{1}(x_1+x_2+x_3)}=\frac{\bell{4}(0)\ell{1}(x_4)}{\bell{1}(x_4)\bell{4}(x_4)\ell{1}'(0)}\left[\tilde\zeta(x_1)+\tilde\zeta(x_2)+\tilde\zeta(x_3)-\tilde\zeta(x_1+x_2+x_3)\right].\label{zeta;sigma;5}
\end{align}
These equations can be proved as follow.  The functions on the left
  and right hand sides are both elliptic functions. According to Liouville's
  theorem two meromorphic functions that have same periods, same zeros and
  poles have constant ratio. If further the two functions coincide at one
  non-trivial point they are identical everywhere. Thus, the corresponding
  equation is proved.

\section{The proof of divergence condition (\ref{eq:divcond1})}\label{Proof;Div}
Introduce two parameters $J_\pm$ as 
\begin{align}
	J_\pm=J_x\pm J_y.
\end{align}
First substitute $a(u+\eta)$ in (\ref{eq:divcond1}) with $a'(u)$. The
divergence condition (\ref{eq:divcond1}) thus implies four identities
\begin{align}
	& 2[a(u)- a'(u)]= J_- \left[ \frac{\bell{4}(u)  \bell{4}(u+\eta)}{  \bell{1}(u)  \bell{1}(u+\eta) }- \frac{\bell{1}(u)  \bell{1}(u+\eta)}{  \bell{4}(u)  \bell{4}(u+\eta) }
	\right], \label{add;1} \\
	& 2[a(u) + a'(u)]=J_+\left[\frac{\bell{4}(u)  \bell{1}(u+\eta)}{\bell{1}(u)\bell{4}(u+\eta)}-  \frac{\bell{1}(u)\bell{4}(u+\eta)}{  \bell{4}(u)  \bell{1}(u+\eta) }   
	\right], \label{aad;2}\\
	& 2 d(u)= J_- \left[\frac{\bell{4}(u)  \bell{4}(u+\eta)}{  \bell{1}(u)  \bell{1}(u+\eta) } + \frac{\bell{1}(u)  \bell{1}(u+\eta)}{  \bell{4}(u)  \bell{4}(u+\eta) }
	\right], \label{add;3}\\
	& 4 J_z- J_+  
	\left[\frac{\bell{4}(u)  \bell{1}(u+\eta)}{  \bell{1}(u)  \bell{4}(u+\eta)} +   \frac{\bell{1}(u)  \bell{4}(u+\eta)}{ \bell{4}(u)  \bell{1}(u+\eta) }   
	\right]+J_-  
	\left[\frac{\bell{4}(u)  \bell{4}(u+\eta)}{  \bell{1}(u)  \bell{1}(u+\eta) } + \frac{\bell{1}(u)  \bell{1}(u+\eta)}{  \bell{4}(u)  \bell{4}(u+\eta) } \right] =0.\label{consistency}
\end{align}
Here we let $u=u_n$.
The above equations can be proved by using elliptic theta function identities. 
With the help of Eqs.~(\ref{theta;11}) and (\ref{theta;44}), we get
\begin{align}
J_-=2\frac{\bell{1}^2(\eta)}{\bell{4}^2(0)},
\qquad J_+=2\frac{\bell{4}^2(\eta)}{\bell{4}^2(0)}.
\end{align}
Then, we have 
\begin{align}
&-J_+ \left[\frac{\bell{4}(u) \bell{1}(u+\eta)}{  \bell{1}(u)  \bell{4}(u+\eta)} +\frac{\bell{1}(u)  \bell{4}(u+\eta)}{\bell{4}(u)\bell{1}(u+\eta)}\right] +J_- \left[\frac{\bell{4}(u)  \bell{4}(u+\eta)}{  \bell{1}(u)  \bell{1}(u+\eta) } + \frac{\bell{1}(u)  \bell{1}(u+\eta)}{  \bell{4}(u)  \bell{4}(u+\eta) } \right]\no\\
&=-2\frac{\bell{4}^2(\eta)}{\bell{4}^2(0)}  
\left[\frac{\bell{4}(u)  \bell{1}(u+\eta)}{  \bell{1}(u)  \bell{4}(u+\eta)} +   \frac{\bell{1}(u)  \bell{4}(u+\eta)}{  \bell{4}(u)  \bell{1}(u+\eta) }   
\right] +2\frac{\bell{1}^2(\eta)}{\bell{4}^2(0)}  
\left[\frac{\bell{1}(u)  \bell{1}(u+\eta)}{  \bell{4}(u) \bell{4}(u+\eta) }+\frac{\bell{4}(u)  \bell{4}(u+\eta)}{  \bell{1}(u)  \bell{1}(u+\eta) } \right]\no\\
&\overset{(\ref{th_th;1})}{=}-2\frac{\bell{1}(u+\eta)\bell{4}(u-\eta)}{\bell{1}(u)\bell{4}(u)}-2\frac{\bell{4}(u+\eta)\bell{1}(u-\eta)}{\bell{1}(u)\bell{4}(u)}\no\\
&\overset{(\ref{theta;14})}{=}-4\frac{\ell{2}(\eta)}{\ell{2}(0)}=-4J_z.	
\end{align}
Thus, the consistency condition (\ref{consistency}) is proved. From Eqs.~(\ref{add;1})-(\ref{consistency}), we get the expression of $a(u)$
\begin{align}
a(u)&=-J_z+\frac{J_+}{2}\frac{\bell{4}(u)\bell{1}(u+\eta)}{\bell{1}(u)\bell{4}(u+\eta)}-\frac{J_-}{2}\frac{\bell{1}(u)\bell{1}(u+\eta)}{\bell{4}(u)\bell{4}(u+\eta)}\no\\
&=-\frac{\ell{2}(\eta)}{\ell{2}(0)}+\frac{\bell{4}^2(\eta)}{\bell{4}^2(0)}\frac{\bell{4}(u)\bell{1}(u+\eta)}{\bell{1}(u)\bell{4}(u+\eta)}-\frac{\bell{1}^2(\eta)}{\bell{4}^2(0)}\frac{\bell{1}(u)\bell{1}(u+\eta)}{\bell{4}(u)\bell{4}(u+\eta)}\no\\
&\overset{(\ref{th_th;2})}{=}-\frac{\ell{2}(\eta)}{\ell{2}(0)}+ \frac{\bell{1}(u+\eta)\bell{4}(u-\eta)}{\bell{1}(u)\bell{4}(u)}\no\\
&\overset{(\ref{theta;14})}{=}\frac{\ell{1}(\eta)\,\ell{2}(u)}{\ell{2}(0)\,\ell{1}(u)}.\label{res:a(u)}
\end{align}
Analogously, we can get $a'(u)\equiv a(u+\eta)$.
The expression of $d(u)$ is obtained as follows
\begin{align}
d(u)&=-J_z + \frac{J_+}{2}  
\left[\frac{\bell{4}(u)\bell{1}(u+\eta)}{\bell{1}(u)  \bell{4}(u+\eta)} + \frac{\bell{1}(u)  \bell{4}(u+\eta)}{\bell{4}(u)  \bell{1}(u+\eta)}  \right]\no\\
&\overset{(\ref{zeta;sigma;1})}{=}\frac{\ell{1}(\eta)}{\ell{1}'(0)}[\zeta(\eta)-2\tilde\zeta(\eta)]+\frac{\bell{4}(\eta)\bell{1}(\eta+\tau)}{\bell{4}(0)\bell{1}(\tau)}\left[\frac{\bell{1}(u+\tau)\bell{1}(u+\eta)}{\bell{1}(u)\bell{1}(u+\eta+\tau)}+\frac{\bell{1}(u+2\tau)\bell{1}(u+\eta+\tau)}{\bell{1}(u+\tau)\bell{1}(u+\eta+2\tau)}\right]\no\\
&\overset{(\ref{zeta;sigma;5})}{=}\frac{\ell{1}(\eta)}{\ell{1}'(0)}[\zeta(\eta)+\tilde\zeta(u)+\tilde\zeta(u+\tau)-\tilde\zeta(u+\eta+\tau)-\tilde\zeta(u+\eta)]\no\\
&\overset{(\ref{zeta;3})}{=}\frac{\ell{1}(\eta)}{\ell{1}'(0)}[\zeta(\eta)+\zeta(u)-\zeta(u+\eta)]=g(\eta)+g(u)+g(u+\eta).\label{res:d(u)}
\end{align}
As noted above, Eq.~(\ref{eq:divcond1}) (or (\ref{eq:divcond2})) is proved analytically. For the bra vector $\phi(u)$, the following equations hold
\begin{align}
	&\phi(u)\otimes\phi(u+\eta)\,h=\phi(u)\otimes\phi(u+\eta)\left[a(u)\sigma^z\otimes \one_2- a(u+\eta) \one_2 \otimes \sigma^z + d(u) \one_4\right],\label{h;phi_1}\\
	&\phi(u)\otimes\phi(u-\eta)\,h=\phi(u)\otimes\phi(u-\eta)\left[-a(u)\sigma^z\otimes \one_2 + a(u-\eta) \one_2 \otimes \sigma^z +  d(-u)\one_4\right],\label{h;phi_2}
\end{align}
where $a(u)$ and $d(u)$ are given by (\ref{a(u,s)}).

\section{Properties of $\psi(u)$ and $\phi(u)$}\label{Proof;ab}

Apart from the divergence conditions, the two-component vectors $\psi(u)$ and
$\phi(u)$ also satisfy the following boundary related equations
\begin{align}
&\vec{h}_1 \vec{\si}_1\,\psi(u)=\left[a_L(u)\sigma_1^z+b_L(u)\right]\psi(u),\label{h1;Ket}\\
&\vec{h}_N \vec{\si}_N\,\psi(u)=\left[a_R(u)\sigma_N^z+b_R(u)\right]\psi(u).\label{hN;Ket}\\
&\phi(u)\vec{h}_1\vec{\sigma}_1=\phi(u)\left[\tilde a_L(u)\sigma_1^z+\tilde b_L(u)\right],\label{phi;1}\\
&\phi(u)\vec{h}_N\vec{\sigma}_N =\phi(u)\left[\tilde a_L(u)\sigma_N^z+\tilde b_R(u)\right],\label{phi;N}
\end{align}
Using Eqs.~(\ref{theta;11})-(\ref{theta;14}), it is easy to get
\begin{align}
&\frac{\bell{4}(u)}{\bell{1}(u)}+\frac{\bell{4}(u)}{\bell{1}(u)}=2\frac{\ell{4}(u)\ell{3}(0)}{\ell{1}(u)\ell{2}(0)},\\
&\frac{\bell{4}(u)}{\bell{1}(u)}-\frac{\bell{4}(u)}{\bell{1}(u)}=2\frac{\ell{4}(0)\ell{3}(u)}{\ell{1}(u)\ell{2}(0)}. 
\end{align}
Then the expressions of $a_L(u)$ and $b_L(u)$ can be obtained directly from
(\ref{h1;Ket})
\begin{align}
a_L(u)&=-\frac{h_x^-}{2}\left[\frac{\bell{4}(u)}{\bell{1}(u)}-\frac{\bell{1}(u)}{\bell{4}(u)}\right]+\frac{i h_y^-}{2}\left[\frac{\bell{4}(u)}{\bell{1}(u)}+\frac{\bell{1}(u)}{\bell{4}(u)}\right]+h_z^-\no\\
&=-\frac{\ell{1}(\eta)\ell{3}(u)}{\ell{1}(u)\ell{2}(0)}\prod_{l=1}^3\frac{\ell{4}(\alpha_l^-)}{\ell{1}(\alpha_l^-)}+\frac{\ell{4}(u)\ell{1}(\eta)}{\ell{1}(u)\ell{2}(0)}\prod_{l=1}^3\frac{\ell{3}(\alpha_l^-)}{\ell{1}(\alpha_l^-)}-\frac{\ell{1}(\eta)}{\ell{2}(0)}\prod_{l=1}^3\frac{\ell{2}(\alpha_l^-)}{\ell{1}(\alpha_l^-)},\no\\	&=-\frac{\ell{1}(\eta)\ell{4}(u-\frac12)}{\ell{1}(u)\ell{2}(0)}\prod_{l=1}^3\frac{\ell{4}(\alpha_l^-)}{\ell{1}(\alpha_l^-)}+\frac{\ell{3}(u-\frac12)\ell{1}(\eta)}{\ell{1}(u)\ell{2}(0)}\prod_{l=1}^3\frac{\ell{3}(\alpha_l^-)}{\ell{1}(\alpha_l^-)}-\frac{\ell{1}(\eta)\ell{2}(u-\frac12)}{\ell{1}(u)\ell{2}(0)}\prod_{l=1}^3\frac{\ell{2}(\alpha_l^-)}{\ell{1}(\alpha_l^-)},\no\\
&\overset{(\ref{additional_formula})}{=}-a(u)+2\frac{\ell{1}(\eta)\prod_{l=0}^3\ell{1}\left(\tfrac{2u+2\chi_l^--1}{4}\right)}{\ell{2}(0)\ell{1}(u)\prod_{k=1}^3\ell{1}(\alpha_k^-)}=a(u)+2\frac{\ell{1}(\eta)\prod_{l=0}^3\ell{1}\left(\tfrac{2u-2\chi_l^--1}{4}\right)}{\ell{2}(0)\ell{1}(u)\prod_{k=1}^3\ell{1}(\alpha_k^-)},\label{aL}\\[4pt]
b_L(u)&=
-\frac{h_x^-}{2}\left[\frac{\bell{4}(u)}{\bell{1}(u)}+\frac{\bell{1}(u)}{\bell{4}(u)}\right]+\frac{i h_y^-}{2}\left[\frac{\bell{4}(u)}{\bell{1}(u)}-\frac{\bell{1}(u)}{\bell{4}(u)}\right]\no\\
&=-\frac{\ell{4}(u)\ell{3}(0)}{\ell{1}(u)\ell{2}(0)}\frac{\ell{1}(\eta)}{\ell{4}(0)}\prod_{l=1}^3\frac{\ell{4}(\alpha_l^-)}{\ell{1}(\alpha_l^-)}+\frac{\ell{3}(u)\ell{4}(0)}{\ell{1}(u)\ell{2}(0)}\frac{\ell{1}(\eta)}{\ell{3}(0)}\prod_{l=1}^3\frac{\ell{3}(\alpha_l^-)}{\ell{1}(\alpha_l^-)},\label{bL}
\end{align}
where
\begin{align}
&\chi_0^\pm=-\alpha_1^\pm-\alpha_2^\pm-\alpha_3^\pm,
\quad \chi_1^\pm=\alpha_2^\pm+\alpha_3^\pm-\alpha_1^\pm,\quad \chi_2^\pm=\alpha_1^\pm-\alpha_2^\pm+\alpha_3^\pm, \quad \chi_3^\pm=\alpha_1^\pm+\alpha_2^\pm-\alpha_3^\pm.\label{chi}
\end{align}
Analogously, we get the expressions for $a_R(u)$ and $b_R(u)$, and for $\tilde
a_k(u),\,\tilde b_k(u),\,k=L,R$
\begin{align}
	\tilde a_L(u)&=-\frac{h_x^-}{2} \left[\frac{\bell{4}(u)}{\bell{1}(u)}-\frac{\bell{1}(u)}{\bell{4}(u)}\right]-\frac{i h_y^-}{2} \left[\frac{\bell{4}(u)}{\bell{1}(u)}+\frac{\bell{1}(u)}{\bell{4}(u)}\right]+h_z^-\no\\
	&\overset{(\ref{additional_formula;2})}{=}-a(u)-2\frac{\ell{1} (\eta ) \prod _{l=0}^3 \ell{4} \left(\frac{2u-2 \chi^- _l+1}{4}\right)}{\ell{2}(0)\ell{1} (u) \prod _{k=1}^3 \ell{1}(\al_k^-)}=a(u)-2\frac{\ell{1} (\eta)\prod _{l=0}^3\ell{4} \left(\frac{2u+2\chi_l^-+1}{4}\right)}{\ell{2}(0)\ell{1}(u)\prod _{k=1}^3 \ell{1}(\al^-_k)},\label{tilde;aL}\\
	\tilde b_L(u)&=-\frac{h_x^-}{2}\left[\frac{\bell{4}(u)}{\bell{1}(u)}+\frac{\bell{1}(u)}{\bell{4}(u)}\right]-\frac{i h_y^-}{2}\left[\frac{\bell{4}(u)}{\bell{1}(u)}-\frac{\bell{1}(u)}{\bell{4}(u)}\right]\no\\
	&=-\frac{\ell{4}(u)\ell{3}(0)}{\ell{1}(u)\ell{2}(0)}\frac{\ell{1}(\eta)}{\ell{4}(0)}\prod_{l=1}^3\frac{\ell{4}(\alpha_l^-)}{\ell{1}(\alpha_l^-)}-\frac{\ell{3}(u)\ell{4}(0)}{\ell{1}(u)\ell{2}(0)}\frac{\ell{1}(\eta)}{\ell{3}(0)}\prod_{l=1}^3\frac{\ell{3}(\alpha_l^-)}{\ell{1}(\alpha_l^-)}\label{tilde;bL}\\
	a_R(u)&=-a_L(u)|_{\al_k^-\to\al_k^+},\qquad  	b_R(u)=-b_L(u)|_{\al_k^-\to\al_k^+},\label{abR}\\
	\tilde a_R(u)&=-\tilde a_L(u)|_{\al_k^-\to\al_k^+},\qquad  	\tilde b_R(u)=-\tilde b_L(u)|_{\al_k^-\to\al_k^+}.\label{tilde;abR}
\end{align}

\section{Proof of the Theorem in Section \ref{sec:Theorem}}\label{Proof;Theorem}
\textbf{Closure}\,\, Like for the $\XXZ$ chain
\cite{PhantomLong,PhantomBetheAnsatz}, $\sigma^z\psi(u)$ can be expanded as
\begin{align}
	\sigma^z\psi(u)&=p_\pm(u)\psi(u)+q_\pm(u)\psi(u\pm2\eta),\label{sigma;psi;1}\\
	\phi(u)\,\sigma^z&=p_\pm(u)\phi(u)+q_\pm(u)\phi(u\pm2\eta).\label{sigma;psi;2}
\end{align}
From (\ref{sigma;psi;1}), we have
\begin{align}
&p_\pm(u)=\frac{\bell{4}(u\pm2\eta)\bell{1}(u)+\bell{1}(u\pm2\eta)\bell{4}(u)}{\bell{4}(u\pm2\eta)\bell{1}(u)-\bell{1}(u\pm2\eta)\bell{4}(u)},\\
&q_\pm(u)=-\frac{2\bell{1}(u)\bell{4}(u)}{\bell{4}(u\pm2\eta)\bell{1}(u)-\bell{1}(u\pm2\eta)\bell{4}(u)}.
\end{align}
With the help of Eqs.~(\ref{Landen})-(\ref{addition;formula}), we simplify the expression of $p_\pm(u)$ and $q_\pm(u)$
\begin{align}
	&p_+(u)=-\frac{\ell{2} (\eta) \ell{1}(u+\eta)}{\ell{1} (\eta) \ell{2}(u+\eta)},\quad q_+(u)=\frac{\ell{2}(0)\ell{1}(u)}{\ell{1}(\eta)\ell{2}(u+\eta)},\label{ab;1}\\
	&p_-(u)=\frac{\ell{2} (\eta) \ell{1}(u-\eta)}{\ell{1} (\eta) \ell{2}(u-\eta)},\quad q_-(u)=-\frac{\ell{2}(0)\ell{1}(u)}{\ell{1}(\eta)\ell{2}(u-\eta)}.\label{ab;2}
\end{align}
From Eqs.~(\ref{aL}), (\ref{tilde;aL}), (\ref{abR}) and (\ref{tilde;abR}), we get the closure condition 
\begin{align}
	&a_L(u_1)=-a(u_1), \qquad a_R(u_{N-2M_+})=a(u_{N-2M_+}),\label{u0;Ket}\\
	&\tilde a_L(v_1)=-\tilde a(v_{1}), \qquad \tilde a_R(v_{N-2M_-})=\tilde a(v_{N-2M_-}).\label{v0;Bra}
\end{align}
Using Eqs.~(\ref{eq:divcond1})-(\ref{eq:divcond2}), (\ref{h1;Ket})-(\ref{hN;Ket}), (\ref{u0;Ket})-(\ref{v0;Bra}) repeatedly, one can prove   
\begin{align}
	H|n_1,\ldots,n_{M_+}\rangle=&\,\mathrm{L.C.}\left(|n_1,\ldots,n_{M_+}\rangle,\,|n_1-1,\ldots,n_{M_+}\rangle,\,|n_1+1,\ldots,n_{M_+}\rangle,\right.\no\\
	&\,\left.\ldots,|n_1,\ldots,n_{M_+}-1\rangle,\,|n_1,\ldots,n_{M_+}+1\rangle\right),\label{Closure;Ket}
\end{align}
where L.C. denotes linear combination and the following terms will not appear on the RHS of (\ref{Closure;Ket})  
\begin{align}
	&|\ldots,n_l,n_{l+1},\ldots\rangle,\quad 1\leq n_l=n_{l+1}\leq N,\no\\
	&|\ldots,n_k,\ldots\rangle, \quad n_k<0, \quad \hbox{or}\quad n_k>N.
\end{align}
Obviously, the ket vectors in (\ref{KetVecs}) form a closed
set. Analogously, the bra vectors form another closed set. The closure
condition (\ref{u0;Ket}) determines the values of $u_1$ and also gives the
constraint in (\ref{SplittingCriterion}).

\textbf{Orthogonality}
We use two sets of parameters $\{\xi_1,\ldots,\xi_N\}$ and $\{\xi'_1,\ldots,\xi'_N\}$ to represent the vectors as
\begin{align}
	\bigotimes_{n=1}^N\psi(u_1+\xi_n\eta),\quad \hbox{and}\quad \bigotimes_{n=1}^N\phi(v_1+\xi'_n\eta).
\end{align}
Obviously, we find
\begin{align}
	\xi_1+\xi'_1
	\leq0,\quad \xi_N+\xi'_N\geq 0,\quad \xi_{n+1}+\xi'_{n+1}-\xi_n-\xi'_n=0,\pm 2,
\end{align}  
So equations $\xi_n+\xi'_n=0$ ($\xi_n+\xi'_n$ is an even number) holds at
least for one point $n\ (1\leq n\leq N)$.  Due to the property
\begin{align}
\phi(u_1+x)\psi(v_1-x)=	\phi(u_1+x)\psi(-u_1-\tau-x)=0,
\end{align}
any pair of vectors $\langle\!\langle n'_1,\ldots,n'_{M_-}|$ and
$|n_1,\ldots,n_{M_+}\rangle$ are mutually orthogonal.

\textbf{Independence} Among the ket vectors in (\ref{Basis;Kets}), there are
only $\sum_{n=0}^{M_+}\binom{N}{n}$ linearly independent basis vectors
\begin{align}
	|0,\ldots,0\rangle, \,\,|0,\ldots,0,n_1\rangle,\,\,|0,\ldots,0,n_1,n_2\rangle,\ldots,|n_1,\ldots,n_{M_+}\rangle,\quad 1\leq n_l\leq N,\quad n_l<n_{l+1}.\label{Basis;Kets;2}
\end{align} 
Consider the ``unfavourable'' case with $\eta=\frac12$. In this case 
\begin{align}
	\phi(u+2\eta)=-\sigma^z\phi(u),\quad \phi(u+4\eta)= \phi(u).
\end{align}
Then, we get
\begin{align}
	&|0,\ldots,0,n_1\rangle\propto\prod_{l=1}^{n_1}\sigma_l^z|0,\ldots,0\rangle,\no\\
	&|0,\ldots,0,n_1,n_2\rangle\propto\prod_{l=n_1+1}^{n_2}\sigma_{l}^z|0,\ldots,0\rangle,\no\\
	&|0,\ldots,0,n_1,n_2,n_3\rangle\propto\prod_{l_1=1}^{n_1}\sigma_{l_1}^z\prod_{l_2=n_2+1}^{n_3}\sigma_{l_2}^z|0,\ldots,0\rangle,\\
	&\qquad\qquad\vdots\no
\end{align}
We see that even in this special setting all the basis vectors in
(\ref{Basis;Kets}) are linearly independent and form an invariant subspace of
the Hamiltonian whose dimension is $\sum_{n=0}^{M_+}\binom{N}{n}$. The bra
vectors in (\ref{BraVecs}) form another invariant subspace of the Hilbert
space whose dimension is
$\sum_{n=0}^{M_-}\binom{N}{n}=2^N-\sum_{n=0}^{M_+}\binom{N}{n}$. The Hilbert
space splits into two invariant subspaces.

\section{The proof of our ansatz for the $M_+=1$ case}\label{Proof;ansatz}

Using Eqs.~(\ref{eq:divcond1}), (\ref{eq:divcond2})-(\ref{hN;Ket}),
(\ref{sigma;psi;1}) repeatedly, we arrive at Eqs.~(\ref{H;n})-(\ref{H;N}) with
\begin{align}
A_+(n)&=\frac{\ell{2}(u_{n-1})}{\ell{2}(u_n)},\qquad
A_-(n)=\frac{\ell{2}(u_{n})}{\ell{2}(u_{n-1})},\label{Apm}\\
X(n)&=2\left[g(u_n+\tfrac12)-g(u_{n-1}+\tfrac12)-2g(\eta)\right],\label{Xn}\\
A_L&=\tfrac12[a_L(u_{-1})+a(u_{-1})]q_+(u_{-1})=\frac{\prod_{l=0}^3\ell{1}\left(\tfrac{2u_{-1}+2\chi_l^--1}{4}\right)}{\ell{2}(u_0)\prod_{k=1}^3\ell{1}(\alpha_k^-)},\label{AL}\\
A_R&=\tfrac12[a_R(u_{N})-a(u_{N})]q_-(u_{N})=\frac{\prod_{l=0}^3\ell{1}\left(\tfrac{2u_N+2\chi_l^+-1}{4}\right)}{\ell{2}(u_{N-1})\prod_{k=1}^3\ell{1}(\alpha_k^+)},\label{AR}\\
X_L&=b_L(u_{-1})-b_L(u_1)+g(u_{-1})-g(u_1)+[a_L(u_{-1})+a(u_{-1})]p_+(u_{-1}),\label{XL}\\
X_R&=b_R(u_{N})-b_R(u_{N-2})-g(u_{N})+g(u_{N-2})+[a_R(u_{N})-a(u_{N})]p_-(u_{N}),\label{XR}\\
E_0(M)&=b_L(u_1)+b_R(u_{N-2M})+(N-1)g(\eta)+g(u_1)-g(u_{N-2M}),\label{E0}
\end{align}
The functional relations in (\ref{recursive;1})-(\ref{recursive;5}) are
fundamental. And Eq.~(\ref{U;property}) is the key of our ansatz. From
Eqs.~(\ref{Y;1}) and (\ref{Xn}), we get the expression for $Y(\l,n)$
\begin{align}
	Y(\pm\l,n)&=g(\l-\tfrac{\eta}{2})-g(\l+\tfrac{\eta}{2})-g(u_n+\tfrac12)+g(u_{n-1}+\tfrac12)+2g(\eta)\no\\
	&=g(\l-\tfrac{\eta}{2})+g(u_{n-1}+\tfrac12)+g(\eta)-g(\l+u_{n}+\tfrac12-\tfrac{\eta}{2})\no\\
	&\quad-g(\l+\tfrac{\eta}{2})-g(u_n+\tfrac12)-g(-\eta)+g(\l+u_n+\tfrac12-\tfrac{\eta}{2})\no\\
	&\overset{(\ref{zeta;sigma;4})}{=}\frac{\ell{1}(\l+\frac{\eta}{2})}{\ell{1}(\l-\frac{\eta}{2})}\frac{\ell{2}(u_{n-1})\ell{2}(\lambda-u_n-\tfrac{\eta}{2})}{\ell{2}(u_n)\ell{2}(\lambda-u_n+\tfrac{\eta}{2})}
	+\frac{\ell{1}(\l-\frac{\eta}{2})}{\ell{1}(\l+\frac{\eta}{2})}\frac{\ell{2}(u_n)\ell{2}(\lambda-u_n+\tfrac{3\eta}{2})}{\ell{2}(u_{n-1})\ell{2}(\lambda-u_n+\tfrac{\eta}{2})}.
\end{align}
It is straightforward to check that $U_n(\l)$ defined in (\ref{def;U})
satisfies Eq.~(\ref{U;property}). Then all our ansatz can be proved
analytically.

\textit{Remark.} Numerical evidence suggests the validity of the following expressions: 
\begin{align}
	\frac{B_-}{B_+}&=\frac{\ell{1}(\l-\al_3^-+\frac{\eta}{2})}{\ell{1}(\l+\al_3^--\frac{\eta}{2})}\prod_{l=1}^2\frac{\ell{1}(\l+\al_l^-+\frac{\eta}{2})}{\ell{1}(\l-\al_l^--\frac{\eta}{2})}\no\\
	&=e^{-8i\pi L_0\l}\left[\frac{\ell{1}(\l+\frac{\eta}{2})}{\ell{1}(\l-\frac{\eta}{2})}\right]^{2N}\prod_{k=1}^3\frac{\ell{1}(\l-\al_k^+-\frac{\eta}{2})}{\ell{1}(\l+\al_k^++\frac{\eta}{2})}.\label{RM}
\end{align}
Based on the above equation, the BAE (\ref{RM}) should have another equivalent form 
\begin{align}
	&e^{-8i\pi L_0\l}\left[\frac{\ell{1}(\l+\frac{\eta}{2})}{\ell{1}(\l-\frac{\eta}{2})}\right]^{2N}\prod_{k=1}^3\frac{\ell{1}(\l-\al_k^+-\frac{\eta}{2})}{\ell{1}(\l+\al_k^++\frac{\eta}{2})}\frac{\ell{1}(\l+\al_3^--\frac{\eta}{2})}{\ell{1}(\l-\al_3^-+\frac{\eta}{2})}\prod_{l=1}^2\frac{\ell{1}(\l-\al_l^--\frac{\eta}{2})}{\ell{1}(\l+\al_l^-+\frac{\eta}{2})}=1.\label{TQ;BAE}
\end{align}
Thus, the BAE (\ref{TQ;BAE}) is consistent with the one given by the other
approach in Ref.~\cite{Yang2006}.

\section{Inhomogeneous BAE for the generic case}
The inhomogeneous BAE of the $\XYZ$ chain with generic open boundary conditions
(pp.192 in \cite{OffDiagonal}) read
\begin{align}
	&\sum_{\sigma=\pm}\sum_{k=1}^3\epsilon_k^\sigma\al_k^\sigma+(N+3+2m)\eta+2\sum_{j=1}^{N+1+m}\mu_j=2l_0\tau+2k_0,\quad \prod_{\sigma=\pm}\prod_{k=1}^3\epsilon_k^\sigma=1,\label{InhoBAE-1}\\
	&c\,e^{-4i\pi l_0(\mu_j+\eta)}\ell{1}^2(\eta)\ell{1}^m(\mu_j)\ell{1}^{2N+m}(\mu_j+\eta)\ell{1}(2\mu_j+\eta)\ell{1}(2\mu_j+2\eta)\no\\
	&=\prod_{\sigma=\pm}\prod_{k=1}^3\frac{\ell{1}(\mu_j+\eps_k^\sigma\al_k^\si+\eta)}{\ell{1}(\eps_k^\sigma\al_k^\si)}\prod_{l=1}^{N+1+m}\ell{1}(\mu_j+\mu_l+\eta)\ell{1}(\mu_j+\mu_l+2\eta),\quad j=1,\ldots,N+1+m,\label{InhoBAE-2}
\end{align}
where $c$ is a constant existing in the inhomogeneous term of the $T$-$Q$
relation and $m=0$ (or $m=1$) for even (odd) $N$. The energy in terms of Bethe
roots is
\begin{align}
	E=-2\sum_{j=1}^{N+1+m}g(\mu_j+\eta)+(N-1)g(\eta)-\sum_{\sigma=\pm}\sum_{l=1}^3g(\eps_l^\si\al_l^\si)-4i\pi l_0\frac{\ell{1}(\eta)}{\ell{1}'(0)}.
\end{align}
When $c=0$, it means that the following types of Bethe roots may exist
\begin{align}
	\mu_j=-\mu_l-\eta, \quad \mu_j=-\mu_l-2\eta,\quad \mu_j=-\eps_k^\si\al_k^\si-\eta.
\end{align}
Suppose that $2m_1$ Bethe roots form pairs as $(\mu_j,-\mu_j+\eta)$,
$2m_2$ Bethe roots form pairs as $(\mu_j,-\mu_j+2\eta)$ and the remaining
$m_s=N+1+m-2m_1-2m_2$ ($m_s$ is an odd number and $1\leq m_s\leq 5$) Bethe
roots are distributed at discrete points $-\eps_k^\si\al_k^\si-\eta$. Then
Eq.~(\ref{InhoBAE-2}) becomes
\begin{align}
	(N-1-2m_1)\eta=\sum_{\sigma=\pm}\sum_{k=1}^3\tilde\epsilon_k^\sigma\al_k^\sigma-2l_0\tau-2k_0,\quad \prod_{\sigma=\pm}\prod_{k=1}^3\tilde\eps_k^\sigma=-1.\label{CBC}
\end{align}

To sum up, if the constraint (\ref{CBC}) is satisfied, the $T$-$Q$ relation and
the BAE degenerate into the homogeneous ones. Due to that the Bethe roots pair
$(\mu,-\mu-2\eta)$ contributes 0 to the energy. Under condition (\ref{CBC}),
the energy becomes
\begin{align}
	E=2\sum_{j=1}^{m_1}[g(\mu_j)-g(\mu_j+\eta)]+(N-1)g(\eta)-\sum_{\sigma=\pm}\sum_{l=1}^3g(\tilde\eps_l^\si\al_l^\si)-4i\pi l_0\frac{\ell{1}(\eta)}{\ell{1}'(0)}.
\end{align}

The degeneration condition (\ref{CBC}) is consistent with our splitting criterion in Eq. (\ref{SplittingCriterion}), which implies the correspondence between homogeneous $T$-$Q$ relations and invariant subspaces.


\end{document}